\documentclass[12pt]{article}

\usepackage{newtxtext,newtxmath}

\usepackage{graphicx}

\usepackage[letterpaper,margin=1in]{geometry}

\renewenvironment{abstract}
	{\quotation}
	{\endquotation}

\date{}

\makeatletter
\renewcommand{\fnum@figure}{\textbf{Figure \thefigure}}
\renewcommand{\fnum@table}{\textbf{Table \thetable}}
\makeatother

\usepackage{scicite}

\usepackage{url}

\providecommand{\num}[1]{\ensuremath{#1}}	

\newcommand{\perspectivearm}{Uprank Bridging}

\newcommand{\perspectivearmdown}{Uprank Bridging, Downrank Toxic}

\newcommand{\surprising}{Challenging Stereotypes}

\newcommand{\diverseapproval}{Diverse Approval}

\newcommand{\addnews}{Add News}

\def\scititle{
	The Prosocial Ranking Challenge: Reducing Polarization on Social Media without Sacrificing Engagement
}
\title{\bfseries \boldmath \scititle}

\author{
	Jonathan~Stray$^{1\ast}$, Ian~Baker$^{1}$, George~Beknazar-Yuzbashev$^{2}$,\and
	Ceren~Budak$^{3}$, Julia~Kamin$^{4}$, Kylan~Rutherford$^{2}$,\and
	Mateusz~Stalinski$^{5}$, Tin~Acosta$^{14}$, Chris~Bail$^{6}$,\and
	Michael~Bernstein$^{7}$, Mark~Brandt$^{8}$, Amy~Bruckman$^{9}$,\and
	Anshuman~Chhabra$^{10}$, Soham~De$^{11}$, Kayla~Duskin$^{12}$,\and
	Sara~Fish$^{13}$, Beth~Goldberg$^{14}$, Andy~Guess$^{15}$,\and
	Dylan~Hadfield-Menell$^{16}$, Muhammed~Haroon$^{17}$, Safwan~Hossain$^{13}$,\and
	Michael~Inzlicht$^{18}$, Gauri~Jain$^{13}$, Zaria~Jalan$^{16}$,\and
	Yanchen~Jiang$^{13}$, Alexander~P.~Landry$^{7}$, Yph~Lelkes$^{19}$,\and
	Hongfan~Lu$^{11}$, Peter~Mason$^{26}$, Jennifer~McCoy$^{20}$,\and
	Smitha~Milli$^{21}$, Paul~Resnick$^{3}$, Emily~Saltz$^{22}$,\and
	Martin~Saveski$^{11}$, Lisa~Schirch$^{23}$, Max~Spohn$^{13}$,\and
	Siddarth~Srinivasan$^{13}$, Alexis~Tatore$^{26}$, Luke~Thorburn$^{24}$,\and
	Joshua~A.~Tucker$^{25}$, Robb~Willer$^{7}$, Magdalena~Wojcieszak$^{17}$,\and
	Manuel~W\"{u}thrich$^{13}$, Sylvan~Zheng$^{25}$
    \\
	\small$^{1}$Center for Human-Compatible AI, UC Berkeley, Berkeley, CA 94720, USA.\and
	\small$^{2}$Columbia University, New York, NY 10027, USA.\and
	\small$^{3}$School of Information, University of Michigan, Ann Arbor, MI 48109, USA.\and
	\small$^{4}$Civic Health Project, USA.\and
	\small$^{5}$University of Warwick, Coventry CV4 7AL, UK.\and
	\small$^{6}$Department of Sociology, Duke University, Durham, NC 27708, USA.\and
	\small$^{7}$Department of Computer Science, Stanford University, Stanford, CA 94305, USA.\and
	\small$^{8}$Department of Psychology, Michigan State University, East Lansing, MI 48824, USA.\and
	\small$^{9}$School of Interactive Computing, Georgia Institute of Technology, Atlanta, GA 30332, USA.\and
	\small$^{10}$University of South Florida, Tampa, FL 33620, USA.\and
	\small$^{11}$University of Washington, Seattle, WA 98195, USA.\and
	\small$^{12}$Information School, University of Washington, Seattle, WA 98195, USA.\and
	\small$^{13}$Harvard University, Cambridge, MA 02138, USA.\and
	\small$^{14}$Jigsaw, Google, New York, NY, USA.\and
	\small$^{15}$Department of Politics, Princeton University, Princeton, NJ 08544, USA.\and
	\small$^{16}$Department of Electrical Engineering and Computer Science, MIT, Cambridge, MA 02139, USA.\and
	\small$^{17}$UC Davis, Davis, CA 95616, USA.\and
	\small$^{18}$Department of Psychology, University of Toronto, Toronto, ON M5S 1A1, Canada.\and
	\small$^{19}$Annenberg School for Communication, University of Pennsylvania, Philadelphia, PA 19104, USA.\and
	\small$^{20}$Department of Political Science, Georgia State University, Atlanta, GA 30303, USA.\and
	\small$^{21}$Meta FAIR, New York, NY, USA.\and
	\small$^{22}$Saltern Studio, USA.\and
	\small$^{23}$Kroc Institute for International Peace Studies, University of Notre Dame, Notre Dame, IN 46556, USA.\and
	\small$^{24}$Department of Informatics, King's College London, London WC2R 2LS, UK.\and
	\small$^{25}$Department of Politics, New York University, New York, NY 10012, USA.\and
    \small$^{26}$Independent.\and
	\small$^\ast$Corresponding author. Email: jonathanstray@berkeley.edu
}

\begin{document} 

\maketitle

\begin{abstract} \bfseries \boldmath
We report the first direct comparisons of multiple alternative social media algorithms on multiple platforms on outcomes of societal interest. We used a browser extension to modify which posts were shown to desktop social media users, randomly assigning 9,386 users to a control group or one of five alternative ranking algorithms which simultaneously altered content across three platforms for six months during the US 2024 presidential election. This reduced our preregistered index of affective polarization by an average of $0.03$ standard deviations ($p < 0.05$), including a $1.5$ degree decrease in differences between the 100 point inparty and outparty feeling thermometers. We saw reductions in active use time for Facebook ($-0.37$ min/day) and Reddit ($-0.2$ min/day), but an \emph{increase} of $0.32$ min/day ($p < 0.01$) for X/Twitter. We  saw an increase in reports of negative social media experiences but found no effects on well-being, news knowledge, outgroup empathy, perceptions of and support for partisan violence. This implies that bridging content can improve some societal outcomes without necessarily conflicting with the engagement-driven business model of social media.
\end{abstract}

\begin{figure}[htbp]
    \centering
    \includegraphics[width=\textwidth]{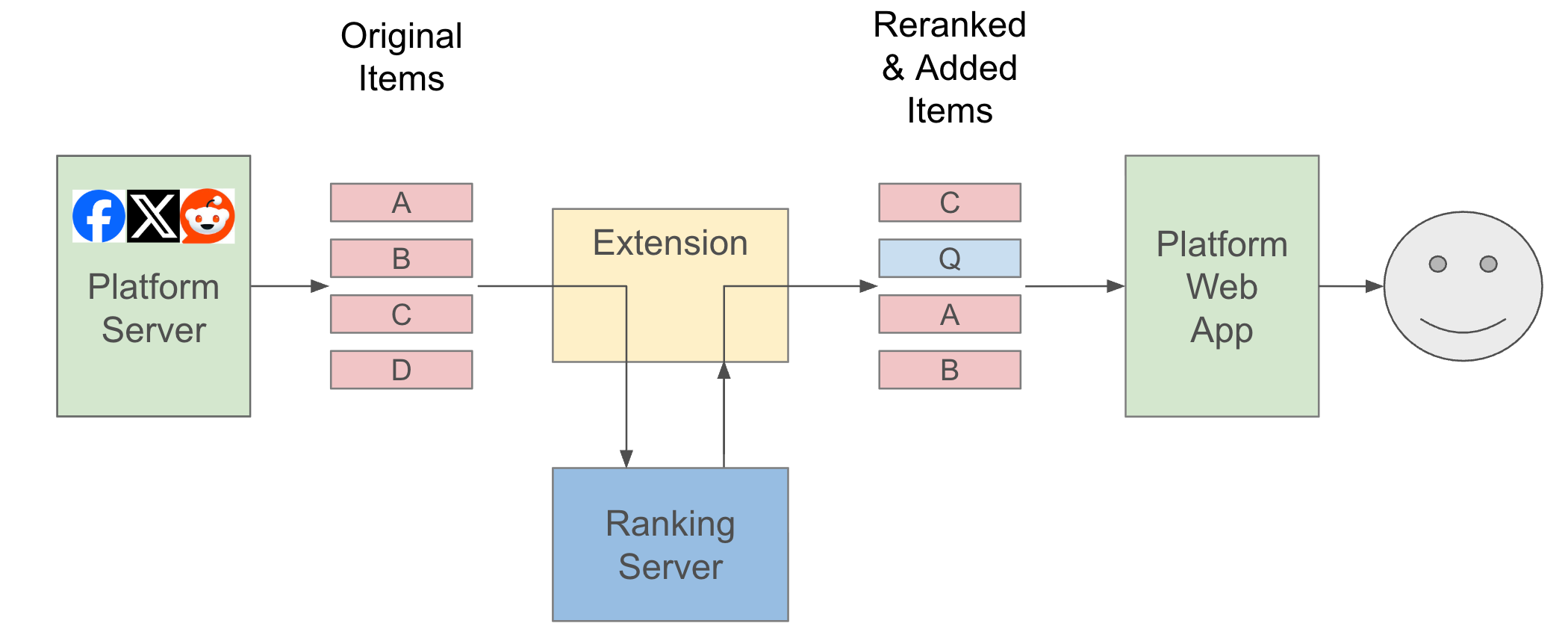} \\
    \includegraphics[width=\textwidth]{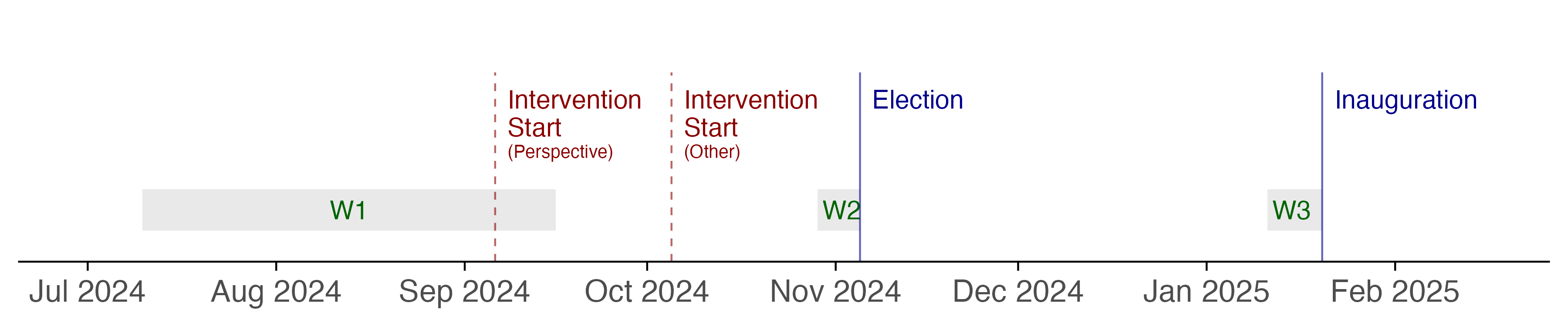}
    \caption{\textbf{Design and timeline of our experiment.} Top: Participants installed a custom browser extension that intercepted platform API calls to extract the top 30-50 items of content served by each of three platforms, rerouting them to our ranking server for real-time reordering, deletion, and addition. Bottom: Experiment timeline, with key dates and survey waves. Wave W1 was the recruitment period, with participants completing a baseline survey on signup. W2 (midline) was pre-election while W3 (endline) was pre-inauguration. \textit{{\diverseapproval}}, \textit{{\surprising}} and \textit{{\addnews}} interventions started later than \textit{{\perspectivearm}} and \textit{{\perspectivearmdown}} due to technical delays. }
    \label{fig:system_diagram}
\end{figure}

\noindent Billions of people around the globe use social media platforms. The backbone of these platforms are ranking algorithms that recommend content largely by choosing items that are predicted to generate views, clicks, watch time, and other engagement \cite{cunningham_ranking_2025}. Critics worry that this selects for identity-confirming, sensational, and polarizing content, potentially exacerbating political polarization \cite{stray_algorithmic_2023}. 

In response to these concerns, researchers and policymakers have called for alternative algorithmic ranking and recommendation strategies that might promote healthier interactions online. Theoretical studies \cite{thorburn_societal_2024}, agent-based simulations \cite{tornberg2023simulating}, observational analyses \cite{fraxanet_analyzing_2025}, leaked platform experiments \cite{thorburn_social_2023}, experiments using simulated social media platforms \cite{jia_embedding_2023}, crowd-sourced misinformation detection systems \cite{wojcik_birdwatch_2022} and expert workshops \cite{goldberg_ai_2024} all suggest that alternative social media algorithms could reduce polarization. Some proposed methods include increasing the diversity of the items users see \cite{stray2022designing}, bridging-based ranking which finds items common to fractured groups \cite{ovadya2021bridging} and ``surprising validators'' where a credible source endorses a stereotypically outgroup idea \cite{glaeser_why_2013}. All of these algorithms are attempts to create impartial mediators that reduce inter-group misperceptions and improve large-scale social relationships. We call such approaches ``prosocial," a term from social psychology that refers to behaviors intentionally designed to benefit others and strengthen social relationships, such as cooperation, empathy, and mutual aid \cite{bateson_altruism_2003}. 

Understanding how such redesigns perform in the field---under real conditions of platform use---is both scientifically important and of high public relevance. Previous large experiments on Facebook and Instagram that minimized reshares \cite{guess2023reshares}, downranked like-minded sources \cite{nyhan2023like}, eliminated political ads \cite{allcott_effects_2025} and re-introduced chronological feeds \cite{guess2023how} found robust null effects on various political attitudes including polarization. However, Piccardi \emph{et al}. \cite{piccardi_reranking_2025} used a browser extension to change the Twitter/X ranking algorithm and showed that a reduction in polarized attitudes is possible. How such results apply across different platforms, for sustained periods, across a vast number of possible interventions is yet to be determined.

In this context, we report a preregistered longitudinal field experiment of multiple prosocial content ranking interventions across multiple platforms, which is both larger (N = 9,386) and longer (six months) than previous tests. We developed three different alternative ranking algorithms through a peer-reviewed competition \cite{stray_prosocial_2024} inspired by the ``megastudy'' approach of Voelkel \emph{et al.} \cite{voelkel2024megastudy}. We solicited entries and had them evaluated by a team of expert judges to identify ranking algorithms which were both practical given our setting, and theoretically-grounded as having the potential for prosocial impacts on users. In addition to these three arms, we selected two more ourselves. \footnote{The Google Jigsaw Perspective API-based treatments \textit{{\perspectivearm}} and \textit{{\perspectivearmdown}} were chosen because they are simple treatments that only re-rank content (not add or delete it), and because similar Perspective-based toxicity classifiers have been widely used in previous scholarship. These experimental arms were funded in part by Google Jigsaw.} Table \ref{tab:rankers} and supplementary text \ref{app:Rankers} present detailed descriptions and theoretical justifications for each experimental arm.
 
We then deployed them as a custom browser extension, extending the technique of Piccardi \textit{et al.} \cite{piccardi2024reranking} to multiple platforms and algorithms. As shown in figure \ref{fig:system_diagram}, this extension modified the ranking of content within algorithmic feeds on Facebook, Reddit, and X/Twitter for 9,386 participants during and after the contentious 2024 U.S. presidential election campaign, from September 2024 to February 2025. Participants were contacted through consumer market research firms and screened for desktop usage of our three social media sites (see supplementary text \ref{app:Recruitment} for details) after which they were asked to install the extension and take the baseline survey via Qualtrics. Only people who kept the extension installed for at least one week are included in our sample.

Each person was randomly assigned to one of five ranking algorithms or control (platform default ranking). Each of these algorithms uses language-model based classifiers to identify and then uprank or add various types of ``prosocial" content, broadly defined as posts or comments with informative, constructive, pro-democracy, or divide-bridging qualities \cite{ovadya2021bridging,jia_embedding_2023}, as follows:

\begin{enumerate}

\item \textbf{{\perspectivearm}}: Rerank posts and comments according to how ``bridging" they are, as evaluated by a suite of experimental classifiers added to Jigsaw's Perspective API \footnote{The Perspective API was originally developed by Google Jigsaw as a ``toxicity" classifier \cite{lees_new_2022} and has become widely used for comment moderation, research, etc. The original toxicity classifier was recently extended to identify more than a dozen different positive and negative communicative constructs \cite{schmer-galunder_annotator_2024,jigsaw_announcing_2024}. For details on the constructs used for our interventions, see supplementary text \ref{app:Rankers}.} \cite{schmer-galunder_annotator_2024,jigsaw_announcing_2024}. Chosen by the authors.

\item \textbf{{\perspectivearmdown}}: Additionally downrank items identified as ``toxic" or outrage-inducing by the same classifier suite. Chosen by the authors.

\item  \textbf{{\surprising}}: Replace political content with ideologically surprising posts that challenge dominant stereotypes of Democrats and Republicans\cite{glaeser_why_2013,dandekar_biased_2013,ozer_partisan_2022}. Chosen through peer-reviewed competition.

\item \textbf{{\diverseapproval}}: Replace political content with posts predicted to be interesting and engaging for both Republicans and Democrats \cite{thorburn_societal_2024,ovadya2021bridging,thorburn_social_2023}. Chosen through peer-reviewed competition.

\item \textbf{{\addnews}}: Add personally relevant news from factual and ideologically diverse news organizations as based on external expert metrics\cite{levy2021social}. Chosen through peer-reviewed competition.

\end{enumerate}

In addition to three waves of surveys, we collected all the content that participating users saw on algorithmic feeds over six months (196 million items) and all the posts and comments they contributed (1.2 million items), as well as all engagement actions on these items including like, share, etc. (84 million events). Given the variety of our treatment arms and nuances in their intended effects, we preregistered examining ten dependent variables, including outcomes relating to polarization, intergroup empathy, news knowledge, mental-health, user experiences, meta-perceptions of support for political violence, and time spent using social media (see supplementary text \ref{app:Outcomes} for details). We preregistered pooling across all treatment arms. In this paper, we focus on a subset of outcomes, with the intention of future papers more closely examining each individual treatment as well as additional dependent variables relevant to each. Pooling across all treatment arms (as preregistered), we found statistically significant, though small, changes on three of these outcomes.

First, we were able to reduce an index of user affective polarization (feeling thermometer and social distance survey questions) by 0.027 standard deviations, including a reduction of about 1.5 degrees on the gap between inparty and outparty evaluations on the 100-point affective polarization scale. This is similar the effect found by Piccardi \emph{et al.} \cite{piccardi_reranking_2025} of 2.1 degrees, and compares with an average increase of 0.6 degrees per year over the past few decades. Second, although we saw reductions in active use time for Facebook ($-0.37$ min/day) and Reddit ($-0.2$ min/day), we found an \emph{increase} of $0.32$ min/day for X/Twitter. Third, we saw a small reduction in an index of reports of positive user experiences. Because participants had been treated for three to four more months by the end of the study while users respond to platform changes over about three months \cite{mladenov_advantage_2019}, we suspect that we measured equilibrium effects. Since these algorithms could be continued indefinitely, these results represent sustainable changes.

Beyond the pooled results, we saw statistically significant polarization reductions for two individual treatments, \emph{{\perspectivearmdown}} and \emph{{\addnews}}. For full results across all outcomes, treatments and platforms see supplementary text \ref{app:Outcomes}.

These results show a complex trade-off between polarization, platform use, and user experiences on social media. We expect that platform implementation would lead to greater polarization reduction due to network effects and creator incentives, and because we could not intervene on mobile apps. Our results have broader implications for platform design and public policy and offer a realistic, scalable and sustainable option for depolarization efforts \cite{holliday_why_2025}. 

\section*{Background}

Disagreement and conflict are a basic feature of democracy and are not necessarily harmful. Problems arise when political disagreement hardens into a form of polarization characterized by strong in-group loyalty and out-group hostility, variously described as affective polarization, political sectarianism, or pernicious polarization \cite{mason2018uncivil,finkel_political_2020,shanto_origins_2019,mccoy_toward_2019}. It increases political intolerance, reduces social trust, weakens support for democratic norms and institutions, foments in-party misinformation endorsement \cite{jenke2024affective}, makes governance and compromise difficult \cite{orhan2022relationship, simonovits2022democratic, kingzette2021affective}, and can lead to political violence \cite{ramsbotham_contemporary_2016}.\footnote{We acknowledge, however, that other scholars find limited evidence for the relationship between affective polarization and support for undemocratic norms or political violence \cite{broockman2023does, druckman2024partisan}.} In the United States, affective polarization between Democrats or liberals and Republicans or conservatives is on the rise \cite{shanto_origins_2019, mason2018uncivil}, and other countries are also experiencing long-term increases in polarization \cite{boxell_cross-country_2020}.

Although various factors contribute to affective polarization, many scholars and observers focus on social media platforms and their recommendation algorithms. Platforms rely largely on user engagement data, i.e., clicks, likes, comments, and shares, to recommend and rank content for users \cite{cunningham_ranking_2025}. Previous work has shown that sensationalist \cite{milli2025engagement}, moralizing \cite{brady_estimating_2025}, ``toxic" \cite{biswas_toxic_2025} and outgroup hostile \cite{rathje_out-group_2021, yu_partisanship_2024} content receives more engagement. Beknazar-Yuzbashev \emph{et al.} \cite{beknazar2022toxic} provided causal evidence for this relationship in the case of toxic content, using a field experiment. By contrast, lab studies where people interact with simulated platforms (e.g. \cite{liu_short-term_2025, kelm_how_2023,jia_embedding_2023}) find inconclusive effects that may not generalize to naturalistic settings of actual platform exposure and interaction. 


Observational studies of real users have been suggestive. Milli \emph{et al.}  \cite{milli2025engagement} found that Twitter users felt angry or sad when viewing political tweets recommended to them, and Oldemburgo de Mello \emph{et al.} \cite{oldemburgo_de_mello_twitter_2024} used experience sampling to determine the correlates of Twitter use, finding short-term decreased subjective well-being and increased affective polarization. To estimate causation, a series of studies in collaboration with Meta altered Facebook and Instagram in various ways such as reverting to a chronological feed \cite{guess2023how}, disabling reshare functions \cite{guess2023reshares} and demoting content from politically congenial sources \cite{nyhan2023like}, all of which had null effects on polarization and other political attitudes. Another recent study by Gauthier \emph{et al.} \cite{gauthier2026political} explored manually switching users between algorithmic and chronological feeds, finding mixed results on engagement and political opinions, but no movement on affective polarization. While these experiments add greatly to our knowledge, many other alternative algorithms, beyond what platforms currently offer, remain to be tested. Recently, Piccardi \emph{et al.} \cite{piccardi_reranking_2025} demonstrated reductions in polarization after ten days of treatment by directly modifying user feeds on Twitter to downrank putatively polarizing content. Our work expands this evidence by testing five novel changes to algorithmic feeds, for six months, on multiple platforms, with a large set of blinded participants.

Last but not least, unlike most work that focuses on minimizing “bad” content on platforms, we increase ``good” or pro-social content in users’ platform feeds.  Given growing evidence shows that platform exposure to and engagement with misleading, untrustworthy, or biased information is in fact relatively rare \cite{altay2022quantifying, budak2024misunderstanding, grinberg2019fake, guess2019less, wojcieszak2022most}, so interventions that aim to decrease exposure to such information may have small effects simply because the majority of users do not encounter it. We instead test a portfolio of pro-social interventions that increase the prevalence of several types of content theorized to defuse polarization \cite{ovadya2021bridging,overgaard_perceiving_2024,voelkel2024megastudy}. 

We built on the wisdom of our research community by launching a challenge, where researchers submitted their ideas for how to redesign social media algorithms. A panel of judges in related academic fields ranked the entries according to scientific interest and experimental plausibility, and our core research team selected five final ``rankers", described in table \ref{tab:rankers}.

\begin{table}
    \centering
    \caption{\textbf{Experimental arms, cohort size, and references giving prior theoretical justification for each intervention.}} 
    \vspace{1em}
    \footnotesize
        \begin{tabular}{|p{1.5cm}|p{7cm}|p{1cm}|p{1cm}|}
            \hline
             \textbf{Name}&  \textbf{Description}&  \textbf{N} & \textbf{Refs}  \\ 
             \hline
             Control group&  Platform default ranking.& \ 2729 &  \\
             \hline
             {\perspectivearm}&  Reorders posts and comments using an average of the experimental bridging attributes in Google Jigsaw’s widely used Perspective API (e.g. compassion, curiosity, reasoning).& 1312 & \cite{schmer-galunder_annotator_2024,jigsaw_announcing_2024}  \\
             \hline
             {\perspectivearmdown}&  Reorders posts and comments using the average of bridging attributes minus the average of negative Perspective API attributes (e.g. insult, identity attack, moral outrage, alienation).& 1400 & \cite{schmer-galunder_annotator_2024,jigsaw_announcing_2024}\\
             \hline
             {\surprising}&  Recommends content classified by GPT-4o as “challenging political stereotypes,” i.e., content in which liberal entities align with stereotypically conservative views or entities, and vice versa. Recommended posts must also pass various quality filters to be shown.& 1391 & \cite{glaeser_why_2013,dandekar_biased_2013,ozer_partisan_2022,voelkel2024megastudy} \\
             \hline
             {\diverseapproval}&  Replaces civic posts (about political or social issues) with posts that are both civic and “bridging,” defined as posts that would be found “valuable, interesting, and engaging” for both average Republicans and Democrats, according to an LLM rater.&  1433 & \cite{thorburn_societal_2024,ovadya2021bridging,thorburn_social_2023}. \\
             \hline
             {\addnews}&  Inserts personalized (topic matched) news posts from 95 credible and ideologically diverse news sources to increase the user’s exposure to factual public affairs information.& 1444 & \cite{levy2021social,haroon2023nudging, askari2024incentivizing} \\
             \hline
        \end{tabular}
    \label{tab:rankers}
\end{table}

\section*{Results}
\subsection*{Experimental Setup}

To examine the effects of our pro-social ranking algorithms, we conducted a large-scale field experiment among consenting participants that combined data from their on-platform behavior with their responses to a multi-wave survey. This design allows us to estimate the effects of the treatment on the information that participants saw, their on-platform behavior, and their political attitudes. In total, the sample consists of 9,386 adult social media users 
who were recruited by multiple survey companies from July to October 2024, with most users recruited in August and September (see supplementary text \ref{app:Recruitment} for details on the recruitment process and sample characteristics). The study received approvals from UC Berkeley IRB and all participants provided informed consent to participate.

Our overall sample was 65\% white, 53\% female, relatively highly educated (43\% with a college degree), and left-leaning (62\% self-identified Democrats and lean-Democrat) (See table \ref{tab:descriptives} in supplementary text \ref{app:Recruitment}). Table \ref{tab:survey_attrition} in supplementary text \ref{app:attrition} shows that there was no differential attrition by arm.

Participants were randomly assigned to one of the five treatments or control. All participants installed a custom browser extension on Facebook, X/Twitter, and Reddit. Each user was assigned to either a control condition, which did not alter the user's feeds, or one of the five algorithmic ``rankers''  that altered what participants saw when viewing algorithmically-ranked feeds of posts or comments (see supplementary text \ref{app:Rankers} for background and implementation details for each intervention). Randomization was successful: the treatment and control groups do not differ in their demographic characteristics (see Table \ref{tab:descriptives} in supplementary text \ref{app:Recruitment}).

Participants in both the treatment and control groups were invited to complete three surveys: one at signup (and therefore at a variable time, 16,823 responses), one in the week before the November 5, 2024 election (5,700 responses), and one in the week before the January 20, 2025 inauguration (3,739 responses). We also injected a single survey question in participants' feeds each week, randomizing between five different questions.

We analyze the final sample of those who completed at least one survey and kept the extension installed for at least one week, in total 9,386 of the 17,095 people who installed the extension. We find no differential attrition: those who stopped using the extension or did not answer the final survey are not statistically significantly different from our final sample (see supplementary text \ref{app:attrition}).

As preregistered, we used a difference-in-differences model for causal identification (see supplementary text \ref{app:Identification} for the full specification). All covariates, namely ideology, gender, race, age and income, were measured in the pre-treatment survey. Supplementary text \ref{app:Outcomes} includes descriptive statistics and balance tables. 

We preregistered ten primary outcomes including polarization, support for partisan violence, perceptions of support for partisan violence, intergroup empathy, platform experiences, mental health, political knowledge, engagement and others (see supplementary text \ref{app:Outcomes} for the definition of all outcomes). All survey questions were taken from prior work. Four of the five rankers were specifically designed to reduce polarization, while the fifth (\emph{{\addnews}}) was designed to increase news knowledge, but a similar field experiment previously also found a polarization reduction \cite{levy2021social}. Three outcomes showed statistically significant effects: polarization, positive user experiences, and active time on platform. We discuss these here and report the remaining preregistered outcomes, plus secondary outcomes and subgroup effects, in supplementary text \ref{app:Outcomes}.

\subsection*{Polarization}
\begin{figure}
    \centering
    \includegraphics[width=\linewidth]{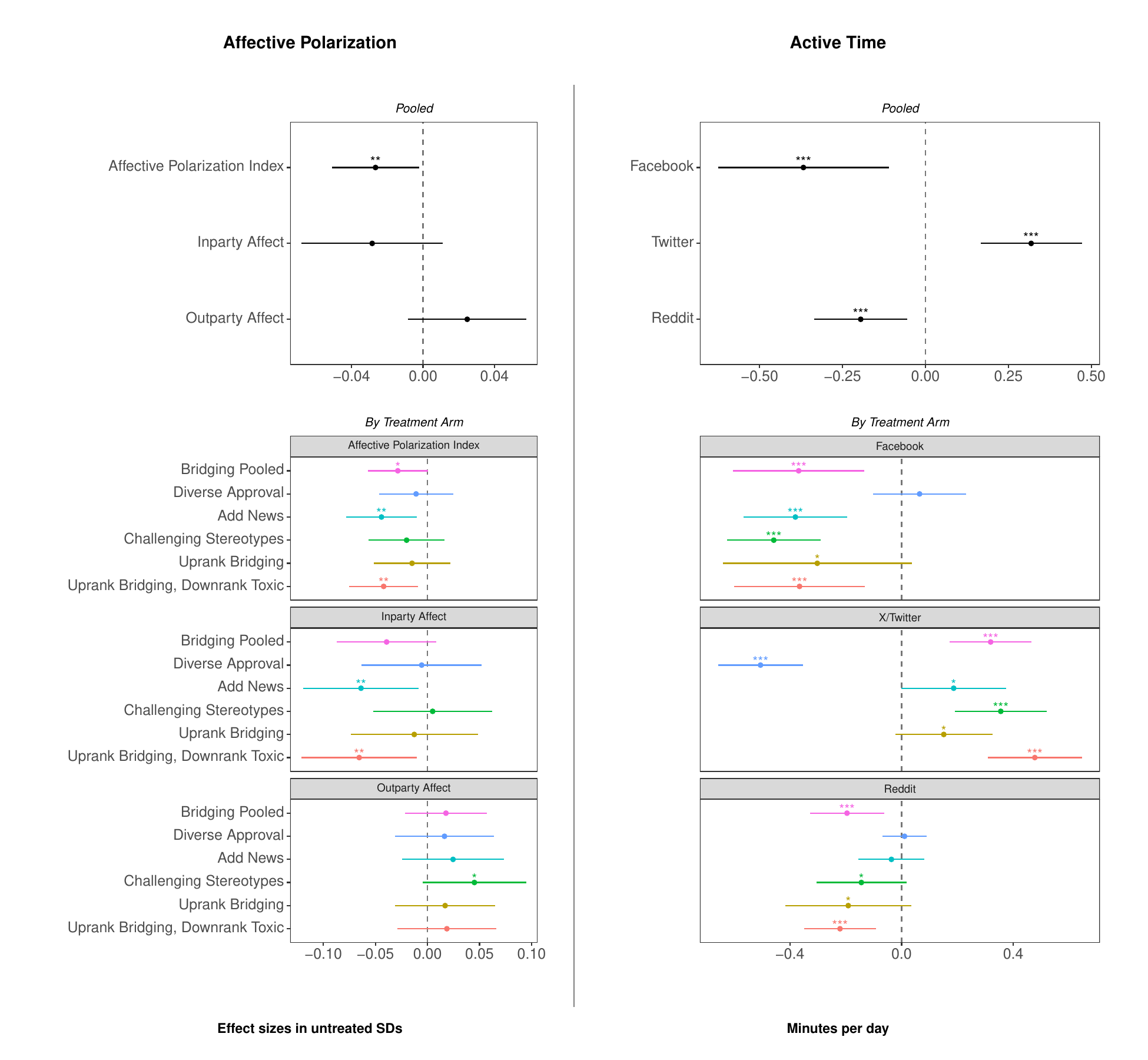}
    \caption{\textbf{Effects on polarization and time on platform.} Left: treatment effects on affective polarization index, and inparty and outparty components of this index, pooled cross all arms and per-arm. All survey outcomes were rescaled to a 0..1 range. Right: treatment effects on active time per platform, pooled across all arms and per-arm. ``Bridging Pooled'' is a preregistered aggregation of {\perspectivearmdown} and {\perspectivearm}, not a separate treatment condition. (* $p<0.1$, ** $p<0.05$, *** $p<0.01$)}
    \label{fig:polarization_and_time}
\end{figure}

Figure \ref{fig:polarization_and_time} shows treatment effects on polarization. As seen, pooled across all experimental arms and platforms we found a reduction in our pre-registered polarization index (combining feeling thermometer ratings and comfort with being friends with counter-partisans). The effect size is 0.027 standard deviations, statistically significant at $p<0.05$. Interestingly, treated participants felt both warmer toward the other side and cooler toward their own side. The result includes a 1.5 point decrease in the gap between inparty and outparty evaluations on the 100 point feeling thermometer scale (see Table \ref{tab:results_survey_components}). Since affective polarization in the U.S. has been rising by about 0.6 points per year in the last four decades, this corresponds to a reversal with a magnitude equal to around two and a half years of average polarization increase. We did not find a significant effect in measures of support for partisan violence or intergroup empathy, the most closely related outcomes. We do, however, find a significant increase in generalized social trust (1.7 points on a 100 point scale, $p<0.05$, Table \ref{tab:results_survey_primary_pooled}). This is consistent with the interpretation that the algorithms improved participants' orientation toward those outside their partisan group.

Figure \ref{fig:polarization_and_time} also plots the effect on polarization index for each ranker. Point estimates for all rankers show reductions in affective polarization, which suggests that a fairly broad set of interventions might work and lends power to our preregistered pooled result. Two of our five arms had statistically significant effects (at $p<0.05$), with all other treatment arms having estimated effects in the same direction. 

Upranking posts and comments that contained nuance, compassion, and personal stories without negative aspects, such as moral outrage, threats, and identity attacks (the \textit{{\perspectivearmdown}} treatment using the experimental Perspective ``bridging API" \cite{schmer-galunder_annotator_2024,jigsaw_announcing_2024}) reduced affective polarization by 0.042 standard deviations. Adding posts from a curated set of verified and ideologically balanced news sources and interleaving them into users' feeds so that up to 15\% of the resulting feed is news (the \textit{{\addnews}} ranker) reduced affective polarization by 0.044 standard deviations.  

\subsection*{Engagement}

Next we examine the impact of our various re-ranking algorithms on platform usage and engagement. Every minute, our extension recorded whether any of our three platforms are the active page on each user's device. We also record every time a tracked content item is newly displayed on the device's view window. Combining these two signals, we calculate daily active time for each user on each platform. The right panels of Figure \ref{fig:polarization_and_time} report our treatment effect on active time across our three treated platforms. In aggregate, we find that our prosocial treatments decrease active time on Facebook and Reddit, but increase time spent on X/Twitter. Looking at the individual classifiers, we find that all treatments except for {\diverseapproval} appear to decrease time spent on Facebook and Reddit, with most having a statistically significant effect. For X/Twitter, all treatments except for {\diverseapproval} appear to increase time on the platform. The {\diverseapproval} treatment has a statistically insignificant effect on both Facebook and Reddit, but significantly decreases time spent on X/Twitter. 

Given that platforms want to maximize time spent on their platforms, we often assume that their algorithms are optimized for active time. Thus, when we alter the ostensibly optimized algorithms, it would make sense that active time would go down. This is exactly what we see on Facebook. However, our results suggest mixed effects for Reddit and the complete opposite effects for X/Twitter. This demonstrates that we cannot assume that every platform is on the Pareto frontier of ``good for society" and ``good for business."

\subsection*{Experience}
We also measured user perceptions of ranking algorithm changes. These survey questions ask the users to recall experiences in the previous two weeks, and tend to respond more readily to algorithmic changes than broader questions about attitudes \cite{cunningham_ranking_2025}. We replicated the questions in the Neely Social Media index \cite{fast_unveiling_2023}, an ongoing survey of social media experiences, combining them into one four-question index. Pooled across all treatment arms we saw an effect of $-0.038$ standard deviations ($p<0.01$) for this Neely index. See supplementary text \ref{app:Outcomes} for survey questions and all per-arm results.

Notably, the \emph{{\perspectivearmdown}} arm did not show statistically significant decreases in the Neely index, while the \emph{{\addnews}} and \emph{{\surprising}} arms had effects of  $-0.068$ ($p<0.01$) and $-0.064$ ($p<0.01$) respectively, driving the aggregate result. We do not know what accounts for the impact seen by these two treatments, but previous work shows that exposure to negative news can lead to negative emotions \cite{shaikh_associations_2024,de_hoog_is_2020} and content which contradicts an individual's preconceptions can be unpleasant. These interventions also replace political content with content from accounts and pages users do not follow, so they might be removing content that users find valuable and meaningful. Hence, these treatments could produce worse experiences for users even if they also reduced polarization.

\section*{Discussion}

Designing algorithms that shift attitudes related to trust, cohesion, and conflict naturally raises serious ethical and political questions. Following modern conflict resolution practice \cite{lederach_little_2014,miall_conflict_2004}, agonistic democratic theory \cite{sax_algorithmic_2022,crawford_can_2016,stray_algorithmic_2023} and previous work considering the connection between social media, polarization, and violence \cite{stray2022designing,stray_algorithmic_2023}, the goal of this work is not to eliminate conflict or enforce consensus, but to transform conflict from destructive to constructive forms. We operationalized this idea by choosing preexisting measures of affective polarization, intergroup empathy, and support for political violence.

The most depolarizing ranking algorithms were \emph{{\perspectivearmdown}} and \emph{{\addnews}}. Reordering content to prioritize constructive attributes alone (\emph{{\perspectivearm}}) did not produce a statistically significant reduction in polarization, whereas a variant that also reduced the prominence of outrage-eliciting content did (\emph{{\perspectivearmdown}}). Although these two treatments were not statistically distinguishable from each other, the results suggest that deprioritizing certain content attributes may be an important aspect of interventions aimed at producing measurable depolarization effects, alongside the promotion of constructive content. Moreover, \emph{Uprank Bridging, Downrank Toxic} also improved generalized social trust. The success of \emph{{\addnews}} demonstrates that downranking content isn't strictly necessary, if it is also possible to inject high-quality content that would not normally appear in feeds. News exposure is not inherently depolarizing, indeed exposure to pro-attitudinal news can increase polarization \cite{garrett_implications_2014,son_ideological_2025}. However, when Levi \textit{et al.}\cite{levy2021social} asked users to manually subscribe to counter-attitudinal news organizations it slightly reduced polarization, with an effect size of $0.03$ standard deviations, comparable to the $0.044$ from our algorithmic change. Surprisingly, we did not find a significant effect on our news knowledge index for this or any other ranker. 

Across all rankers, our experiment produced a mean 1.5 point shift the 100-point feeling thermometer scale. The most similar previous experiments are Piccardi \textit{et al.} \cite{piccardi_reranking_2025} which produced polarization decreases corresponding to 2.1 points, and Levi \textit{et al.} \cite{levy2021social} which reduced polarization by 0.96 points. This suggests that this line of research may be producing replicable results.

A recent meta-analysis of depolarization interventions found a mean 5.4 point shift on the 100-point feeling thermometer scale \cite{holliday_why_2025}. That review raised significant concerns about the durability and scalability of the depolarization techniques studied, which were mostly ``one-shot" interactions. While our results are smaller they are also sustainable and scalable, because an algorithmic change can be applied simultaneously to all users of a platform and run indefinitely. Industry experience suggests that user adjustment to algorithmic changes generally happens over three months or so \cite{mladenov_advantage_2019}, so our six month experiment likely captured equilibrium effects.

Many authors have argued that meaningful reductions in polarized attitudes and behaviors can only come from ``structural" changes \cite{wang_systems_2021,coleman_taking_2019,voelkel2024megastudy} which alter the rules or incentives for politicians, journalists, voters, etc. Platform algorithm changes are just such a structural change. Our experiment estimates the first-order effects of exposure to different content, but our design cannot estimate the second-order effects that would result from applying such changes to all users of a platform \cite{eckles_estimating_2016}. Related experiments \cite{brady2021how} show that users who are less polarized in turn like and share less polarizing content, so these network effects are expected to be positive. We were also unable to change people's feeds on mobile devices where most social media consumption happens, further biasing our effect downwards. Creator incentives are potentially even more significant, and algorithmically-induced creator behavior shifts have been both theorized \cite{hu_incentivizing_2024} and observed \cite{garz_algorithmic_2023,stray_algorithmic_2023}. If widely-used ranking algorithms prioritize more constructive content, then major content creators (including journalists and politicians) may change their behavior accordingly. 

Nevertheless, modifications to platform algorithms alone will not solve longstanding polarization. Escaping from the escalating conflict dynamics seen in many modern democracies will require changes across all of society \cite{burgess_applying_2022}. Yet social media algorithms have often been suggested as a key intervention point with potentially far-reaching consequences. By testing across multiple algorithms and platforms, this work provides robust evidence that it is indeed possible to shift the polarization dynamics of social media, at least a little and maybe more. Further, we found the effects on engagement to be small and sometimes positive, which suggests that economic tradeoffs do not preclude healthier algorithms. Finally, our study also reveals that reductions in affective polarization might come at a cost of negative impacts in other meaningful outcomes, such as user experiences for our participants. As such, it is advisable to monitor the effects of interventions for potential unintended negative consequences.


\clearpage 

%
\bibliography{references} 
\bibliographystyle{sciencemag}

%
%
%
%
%
%


\section*{Acknowledgments}
We would like to thank our contest judges (some of whom are also authors): Jennifer McCoy, Robb Willer, Alexander P. Landry, Amy Bruckman, Andy Guess, Chris Bail, Dylan Hadfield-Menell, Josh Tucker, Lisa Schirch, Magdalena Wojcieszak, Mark Brandt, Michael Bernstein, Michael Inzlicht, Paul Resnick, and Yph Lelkes. This research would not have been possible without our tireless engineering staff who created our complex software on tight deadlines and kept it running: Leo Alekseyev, Brian Zimmer, Sana Pandey, James Clark, Cameron Yuen, Haroon Iftikhar, Amar Humackic, Harun Skender, Ante Sunjic, Nedzid Zaklan, and Jayson Vantuyl. Elizbeth Cooper and Sawyer Bernath at the Berkeley Existential Risk Initiative provided much-needed operational and administrative support. Finally, thank you to the Civic Health Project for bringing the core team together.

\paragraph*{Funding:}
Funding for this experiment was provided by the DALHAP Foundation, Google Jigsaw, Survival and Flourishing Fund, Project Liberty, the Harvard Berkman-Klein Center, Dan McCormick, Reset.Tech, and Justin Kosselyn. J.K. was supported by the Civic Health Project. S.D., K.D., and M.S. were supported by National Science Foundation Award IIS-2403434.

\paragraph*{Author Contributions:}
J.S. conceived the study, raised funds, recruited participants, and made final editorial decisions. J.S., G.B.Y., C.B., J.K., K.R. and M.S. designed the study and wrote the paper. I.B. led engineering and made related scientific decisions. G.B.Y., M.S., J.S., and S.Z. performed data analysis. M.Wo. contributed additional writing. B.G., T.A. and E.S. implemented the Perspective API bridging classifiers. L.T., S.D., K.D., H.L., S.M. and M.S. implemented the {\diverseapproval} ranking algorithm. S.S., M.S., P.M., G.J., Y.J., S.F., A.T., S.H., and M.Wü. implemented the {\surprising} ranking algorithm. M.M., A.C., J.S., and M.Wo. implemented the {\addnews} ranking algorithm. J.M., R.W., A.P.L, A.G., C.B., D.H.M., J.T., L.S., M.Wo., M.Be., M.Br., M.I., P.R. and Y.L. judged the contest.

\paragraph*{Competing interests:}
``There are no competing interests to declare.''

\paragraph*{Data and materials availability:}
TBD

\newpage


\renewcommand{\thefigure}{S\arabic{figure}}
\renewcommand{\thetable}{S\arabic{table}}
\renewcommand{\theequation}{S\arabic{equation}}
\renewcommand{\thepage}{S\arabic{page}}
\setcounter{figure}{0}
\setcounter{table}{0}
\setcounter{equation}{0}
\setcounter{page}{1} 


\begin{center}
\section*{Supplementary Materials for\\ \scititle}

Jonathan~Stray$^{\ast}$, Ian~Baker, George~Beknazar-Yuzbashev,\\
Ceren~Budak, Julia~Kamin, Kylan~Rutherford,\\
Mateusz~Stalinski, Tin~Acosta, Chris~Bail,\\
Michael~Bernstein, Mark~Brandt, Amy~Bruckman,\\
Anshuman~Chhabra, Soham~De, Kayla~Duskin,\\
Sara~Fish, Beth~Goldberg, Andy~Guess,\\
Dylan~Hadfield-Menell, Muhammed~Haroon, Safwan~Hossain,\\
Michael~Inzlicht, Gauri~Jain, Yanchen~Jiang,\\
Alexander~P.~Landry, Yph~Lelkes, Hongfan~Lu,\\
Peter~Mason, Jennifer~McCoy, Smitha~Milli,\\
Paul~Resnick, Emily~Saltz, Martin~Saveski,\\
Lisa~Schirch, Max~Spohn, Siddarth~Srinivasan,\\
Alexis~Tatore, Luke~Thorburn, Joshua~A.~Tucker,\\
Robb~Willer, Magdalena~Wojcieszak, Manuel~W\"{u}thrich,\\
Sylvan~Zheng\\
\small$^\ast$Corresponding author. Email: jonathanstray@berkeley.edu
\end{center}

\newpage


\setcounter{section}{0}
\section{Supplementary Materials}

We built a custom Chrome browser extension that intercepts communication between the web client and platform server for Facebook, X/Twitter, and Reddit, building on previous browser extension platform manipulation approaches \cite{piccardi2024reranking,beknazar2022toxic}. When the application page loads, our extension immediately queries the platform server for the first 50-100 posts or comments (depending on platform, see supplementary text \ref{app:Rankers}), then sends the text content of these posts to our own server which runs all algorithmic interventions. Each tested ``ranker" can reorder, remove, or add content. This extends previous browser extension-based methods which could not add content. The ranker forwards the altered feed back to the web client, where it is displayed to the user. Users in the control condition go through the same server with a pass-thru ranker. All results (experimental and control) are returned by the server after a fixed 500ms, to prevent latency differences from contaminating the comparison (see supplementary text \ref{app:Rankers}).

We recruited users through a combination of research panel providers and consumer market research companies, screening for people who reported using at least one of Facebook, X/Twitter or Reddit and at least 30\% of their social media use on desktop. 17,095 users installed the browser extension within our enrollment period from July to September 2024 of which 16,823 completed the baseline survey, but many soon removed the extension, leaving 9,386 who kept it installed for at least one week. We administered the midline survey from October 29 to November 5, recording 5,700 responses. We administered the endline survey from January 10 to January 20, recording 3,739 responses. 5,294 people still had the extension installed on January 10, so we were still able to collect engagement and behavioral data on many people who did not fill out the endline survey.

We had five different experimental arms each corresponding to one ranking algorithm. We randomly assigned participants at enrollment to one arm (1/7 of all users per arm) or control (2/7 of all users) for the entire experiment. Due to technical challenges, different experimental arms started interventions at different dates (Early September to early October), but all users were held in the control condition (no intervention) for at least 14 days post sign-up to record baseline behavioral data. 

Effects were estimated using a difference-in-differences model as detailed in supplementary text \ref{app:Identification}. Because even within the same study arm different users started treatment on different dates, we used additional fixed effects to account for this variation. Regardless of intervention start date, all users of all arms were surveyed during the same week-long midline and endline periods.


\subsection{Preregistration and Deviations}

We preregistered our study in the AEA RCT Registry on August 29, 2024, prior to the start of interventions.\footnote{The link to the preregistration: \url{https://www.socialscienceregistry.org/trials/14274}.} The initial registration covered Treatment Arms 1--2. Treatment Arms 3--5 were added via an amendment submitted prior to the start of those interventions.

We did our best to ensure that our experiment and subsequent analyses followed what was preregistered. Below, we outline deviations that occurred.

\subsubsection{Outcome Measures}

Our preregistration listed ``meaningful connections on platforms'' and ``experiences on platforms'' as separate primary outcomes (outcomes 7 and 8). In practice, the survey questions for both outcomes are drawn from the Neely Social Media Index \cite{fast_unveiling_2023}, which was designed as a unified instrument measuring positive and negative user experiences on social media platforms. We therefore report these as a single combined index (the ``Neely index''), consistent with the instrument's intended use.

For affective polarization, following the standard operationalization \cite{iyengar_affect_2012}, we computed the difference between inparty and outparty scores for both the feeling thermometer and the friendship comfort questions, then averaged these two difference scores. The preregistration does not specify this precisely.

Our preregistration mistakenly specified that we will measure social trust using both a general trust question and one of our affective polarization questions about being comfortable having friends from the outparty. We should have only included the first question. To address this, we report treatment effects on general trust alone (see \textit{Trust} columns in Table \ref{tab:results_survey_primary} and Table \ref{tab:results_survey_primary_pooled}) and on the two question index, as preregistered, (see \textit{TrustIndex} columns in Table \ref{tab:results_survey_primary} and Table \ref{tab:results_survey_primary_pooled}). Both outcomes show significant treatment effects of the pooled algorithms against the control, making the preregistration deviation inconsequential.

Our preregistration specified engagement rate (total engagement signals divided by number of posts seen) as a primary measure of social media use alongside active time. A software issue compromised our ability to record view events on Facebook before September, rendering the post-view denominator unreliable for this measure. We therefore report total engagement events as an alternative measure of engagement (see Appendix~\ref{sec:engagement}).

Lastly, other engagement-based secondary outcomes, including total number of political/civic posts seen, average toxicity of posts seen, engagement rate with toxicity, engagement rate with political/civic posts, average toxicity of posts created, will be reported in a companion study. The same applies to heterogeneity analysis with respect to measures dependent on these outcomes, such as treatment effect heterogeneity by baseline exposure to political or toxic content.

\subsubsection{Sample Size}

We recruited 16{,}823 participants who installed the browser
extension and completed the baseline survey, exceeding the
preregistered target of approximately 15{,}000. However, many participants did not use the extension for more than a few days, leaving 9{,}386 who kept it active for at least one week---a threshold that ensures some level of authentic user activity and a minimum period of baseline behavioral data for our difference-in-differences estimation. Of these, 3{,}739 completed the endline survey, well below the upper bound of 80\% retention (approximately 12{,}000) specified in the preregistration. Section \ref{app:attrition} shows no evidence of differential attrition by treatment arm. Section \ref{app:robustness_inclusion} shows that our results are robust to changing the inclusion requirement to two weeks or two days.

\subsubsection{Interventions}

The preregistration specified re-ranking the first 50 content items per page load. In practice, slate sizes varied by platform and content type (see Table \ref{tab:slate_length}). Most notably, Facebook posts were fetched in two passes (10, then 35) because Facebook returns posts in small batches, and waiting for 50 before displaying any content would have caused unacceptable delays for users.

We removed the \textit{nuance} classifier from the \textit{\perspectivearm} and \textit{\perspectivearmdown} interventions because it ended up mostly picking up on comment length, i.e. it did not work as intended because of poor training data. From \textit{\perspectivearmdown} we removed \textit{severe toxicity} because it turned out to be nearly identical to \textit{toxicity}, removed \textit{profanity}, \textit{sexually explicit} and \textit{flirtation} because they exhibited too many false positives on actually constructive content, and removed \textit{power appeal} because it was never implemented.

\subsection{Results on All Outcomes}\label{app:Outcomes}

Here we present tabular results on all of our preregistered outcomes.
Our main results are aggregate effects across all experimental arms. This analytical strategy addresses the following research question: can modifications to existing social media algorithms designed by informed researchers reduce polarization?
Consistent with our preregistration plan, we also report effects individually by ranker arm to facilitate comparisons between different ranking algorithms, as well as 
heterogeneous effects by platform (where applicable), and demographic variables (race, gender, and partisanship).
We additionally preregistered examining results by baseline levels of political content and toxic content exposure - heterogeneous effects by these variables are omitted in this draft as they are currently under construction. 

All analysis models for survey based outcomes use standardized survey indices such that effects are measured in standard deviations of the control group. 
Platform based outcomes are reported in raw units (minutes per day, for active time). 
For a complete description of the identification strategy and analysis model specification details, see supplementary text \ref{app:Identification}. 

\subsection{Survey Outcomes}

Our preregistered survey outcomes included \begin{enumerate}
    \item mental health and well being \cite{topp_who-5_2015}
    \item support for political violence \cite{kalmoe_lethal_2019}
    \item perceptions of the outgroup's support for political violence (aka "meta-perceptions") \cite{mernyk_correcting_2022}
    \item affective polarization \cite{iyengar_affect_2012}
    \item inter-group empathy \cite{sirin_group_2016}
    \item political knowledge 
    \item meaningful connections and experiences on social media \cite{fast_unveiling_2023}
    \item negative emotions
    \item social trust 
\end{enumerate}

Outcomes 1-8 were preregistered as primary outcomes, while outcome 9 was registered as a secondary outcome. All outcomes were preregistered both pooled across all treatment arms and for individual arms.

\subsection{Preregistered Indexes}

For most of our survey outcomes, we asked a combination of related survey questions. We preregistered the following approaches to creating indexes for these outcomes. Unless otherwise specified, we standardize scores for each statement or question, and add them up to create a single index measure for each outcome.

\paragraph{Mental Health}
Mental Health was measured using the following survey questions (All of the time, Most of the time, More than half the time, Some of the time, At no time) from the WHO-5 well-being index \cite{topp_who-5_2015}:
\begin{itemize}
    \item I have felt cheerful and in good spirits
    \item I have felt calm and relaxed
    \item I have felt active and vigorous
    \item I woke up feeling fresh and rested
    \item My daily life has been filled with things that interest me
\end{itemize}

\paragraph{Support for Partisan Violence}

Support for partisan violence was measured using the following survey questions from \cite{kalmoe_lethal_2019}:
\begin{itemize}
    \item When, if ever, is it OK for [inparty] to send threatening and intimidating messages to [outparty] leaders? (101 pt scale, Never, Always)
    \item How much do you feel it is justified for [inparty members] to use violence in advancing their political goals these days? (101 pt scale, Not at all, Extremely)
\end{itemize}

\paragraph{Meta-perceptions on Support for Partisan Violence}

Meta-perceptions on support for partisan violence was measured using two survey questions (All of the time, Most of the time, More than half the time, Some of the time, At no time) from \cite{mernyk_correcting_2022}:

\begin{itemize}
    \item How do you think an average [outparty] would respond to the following question? How much do you feel it is justified for [outparty] to use violence in advancing their political goals these days?
    \item How do you think an average [outparty] would respond to the following question? When, if ever, is it OK for [outparty] to send threatening and intimidating messages to [inparty] leaders?
\end{itemize}

\paragraph{Affective Polarization}

Affective polarization was measured using the following survey questions, slight variations of the ANES survey questions asked for many years \cite{iyengar_affect_2012}:
 
\begin{itemize}
    \item Please indicate how you feel toward [outparty members] using the scale below. 100 means that you feel very favorably or warm toward them, 0 that you feel very unfavorable or cold, and 50 are neutral.
    \item Please indicate how you feel toward [inparty members] using the scale below. 100 means that you feel very favorably or warm toward them, 0 that you feel very unfavorable or cold, and 50 are neutral.
    \item How comfortable are you having friends who are [outgroup members]? (1-7 scale from Not at all to Extremely)
    \item How comfortable are you having friends who are [ingroup members]? (1-7 scale from Not at all to Extremely)
\end{itemize}

To aggregate these measures, we first standardized scores for each statement. Then, we took the average of the inparty and the negative of outparty scores for both the feeling thermometer questions and questions about having friends of each party. We then took the average of those two difference scores.

\paragraph{Intergroup empathy}

Intergroup empathy was measured using two survey questions (7-point scale ranging from Strongly disagree to Strongly agree) from \cite{sirin_group_2016}:

\begin{itemize}
    \item I find it difficult to see things from [outparty] point of view.
    \item I think It is important to understand [outparty] by imagining how things look from their perspective.
\end{itemize}

\paragraph{Political Knowledge}

Participants were asked questions about their political knowledge in each of the surveys. In each survey, they were given five news headlines and asked, ``Of the following news events, which ones do you think are true events that occurred in the last month, and which ones do you think are false and did not occur?'' (True, False, Unsure)

For each survey, we gathered recent headlines using archives of the Google News page from up to 2 weeks before each survey, selected five which were recent, among the top stories, and relevant to politics, and then changed two headlines to be false versions.

\paragraph{Meaningful Connections and Experience}

Meaningful connection and experiences on platforms was measured using the following survey questions, from the Neely Social Media Index survey \cite{fast_unveiling_2023}:

\begin{itemize}
    \item In the last two weeks, have you experienced a meaningful connection with others on Facebook?
    \item In the last two weeks, have you experienced a meaningful connection with others on X (Twitter)?
    \item In the last two weeks, have you experienced a meaningful connection with others on Reddit?
    \item In the last two weeks, have you personally witnessed or experienced something that affected you negatively on Facebook?
    \item In the last two weeks, have you personally witnessed or experienced something that affected you negatively on X (Twitter)?
    \item In the last two weeks, have you personally witnessed or experienced something that affected you negatively on Reddit?
    \item In the last two weeks, have you learned something that was useful or helped you understand something important on Facebook?
    \item In the last two weeks, have you learned something that was useful or helped you understand something important on X (Twitter)?
    \item In the last two weeks, have you learned something that was useful or helped you understand something important on Reddit?
    \item In the last two weeks, have you witnessed or experienced content that you would consider bad for the world on Facebook?
    \item In the last two weeks, have you witnessed or experienced content that you would consider bad for the world on X (Twitter)?
    \item In the last two weeks, have you witnessed or experienced content that you would consider bad for the world on Reddit?
\end{itemize}

\paragraph{Negative Emotions}
We asked about negative emotions only on the in-feed survey, see supplementary text \ref{app:in-feed}.

\paragraph{Social Trust}
This question is from the General Social Survey, answered on a scale of 0 (Most people can be trusted) to 100 (You can't be too careful).

\begin{itemize}
    \item 
Generally speaking, would you say that most people can be trusted, or that you can't be too careful in dealing with people?
\end{itemize}

\subsection{Engagement Outcomes}
\label{sec:engagement}

Our primary measure of engagement (reported in the main text) was total active time spent on platform. We calculated this using the methodology described in \cite{beknazar2022toxic}. 
Our other preregistered measures of engagement include engagement rate, total number of posts seen,  total number of political/civic posts seen average toxicity of posts seen, toxic post engagement rate, political/civic post engagement rate, and average toxicity of created posts. 
However, a software problem compromised our ability to record view events on Facebook before September. Thus, we are unable to calculate the secondary preregistered engagement measures which depend on view count integrity. Instead, we present raw number of engagement events, which we believe serves as a reliable secondary measure of engagement. 

Treatment effects on engagement outcomes are reported as aggregate effects over all treatment arms as well as individually across each ranker arm. We also present heterogeneous effects by party, gender, and race. All results presented both combined across platforms and disaggregated.

\section{In Feed Survey Measures}
\label{app:in-feed}

\begin{figure}
    \centering
    \includegraphics[scale=0.2]{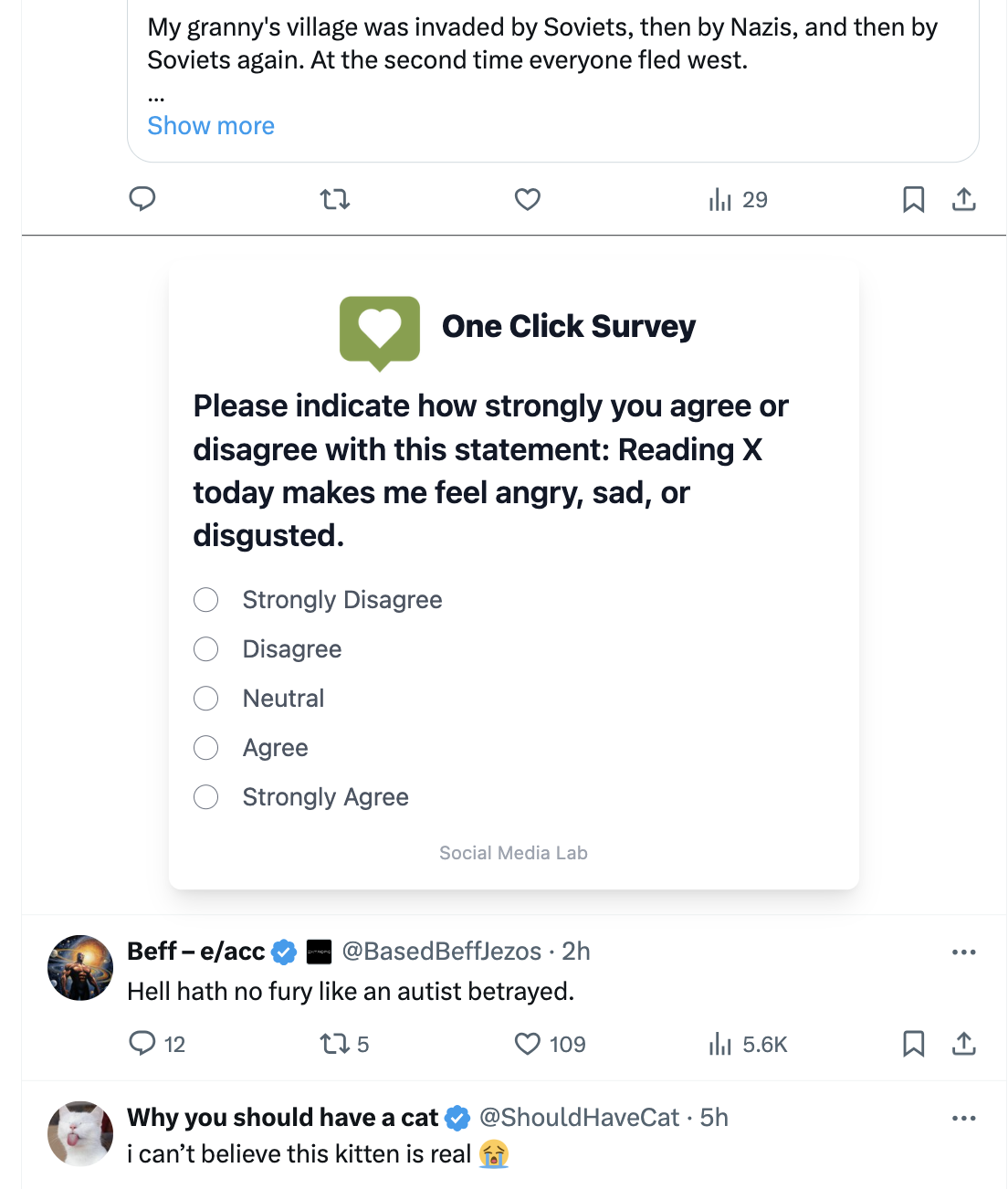}
    \caption{\textbf{How in-feed surveys appeared to participants on X/Twitter.} Similar UI was used for Facebook and Reddit.}
    \label{fig:in_feed_survey}
\end{figure}

In addition to our three full surveys sent to participants at baseline, midline, and endline, we also periodically sent a small set of survey questions to participants, embedded in their feeds as in figure \ref{fig:in_feed_survey}. Each survey appeared one week after the most recent previous appearance, and was shown on whatever platform the user used first, up to three times on three consecutive days. We preregistered reporting results for these questions. We were hoping to measure short-term effects with these in-feed surveys, akin the experience sampling method of \cite{oldemburgo_de_mello_twitter_2024} which measured affect within 30 minutes of Twitter usage. The results are shown in table \ref{tab:in_feed}, estimated with the two-way fixed effects model of section \ref{app:Identification}. We found no statistically significant effects for these outcomes by treatment, possibly because of the relatively low response rate. Note that the N in table reports the number of survey answers (users times weeks), not the number of individual users who answered.

The following questions were asked in-feed:

\begin{itemize}
    \item  Please indicate how you feel toward [out-party] using the scale below [10 - Very favorable (warm toward them), 9, 8, 7, 6, 5 - Neutral, 4, 3, 2, 1, 0 - Very unfavorable (cold toward them)]

    \item How interested would you say you have been in the political campaigns in the last two weeks? [Not at all interested, A little interested, Somewhat interested, Very interested, Extremely interested] (Asked before Nov 5th 2024)

    \item How interested would you say you have been in politics in the last two weeks? [Not at all interested, A little interested, Somewhat interested, Very interested, Extremely interested] (Asked after Nov 5th 2024)

    \item Please indicate how strongly you agree or disagree with this statement: Reading [current platform] today makes me feel angry, sad, or disgusted. [Strongly Disagree, Disagree, Neutral, Agree, Strongly Agree]

    \item How do you think an average [out-party] would respond to the following question: How much do you feel it is justified for [out-party] to use violence in advancing their political goals these days? [0 - Not at all, 1, 2, 3, 4, 5, 6, 7, 8, 9, 10 - Extremely]

    \item Think about the coming presidential election. If [inparty candidate] is declared the winner of a contested election, how likely do you think [out-party] voters would be to engage in violence? [0 - Not at all likely, 1, 2, 3, 4, 5, 6, 7, 8, 9 ,10 - Extremely likely] (Asked before Nov 5th 2024)

    \item Please indicate how strongly you agree or disagree with this statement: I find it difficult to see things from [out-party] point of view [Strongly Disagree, Disagree, Neutral, Agree, Strongly Agree"]
\end{itemize}

\section{Ranking Algorithms}
\label{app:Rankers}

As shown in figure \ref{fig:system_diagram}, our browser extension intercepted API calls from the platform web client code and sent them to our servers for modification. To reduce latency, only the first N posts were retrieved from the platform server and sent for re-ranking. This slate length N varied by platform, and also depending on whether the user was viewing posts, comments, or replies. Our extension did not send further posts for re-ranking after this first N, because it would not be possible to re-order posts from the second batch into the first or vice-versa. The exception was Facebook posts, where we fetched first 10 and then another 35 in the background. This was necessary because Facebook only returns posts in groups of 5, so fetching 50 Facebook posts before displaying any posts at all would cause unacceptably long delays. See table \ref{tab:slate_length} for the full logic. 

In each slate of posts, ads and items without text content (videos and photos without captions) were treated as ``immovable" and left in place, while other posts were re-ordered around them. When available, photo and video alt text was appended to the main text, and on some platforms (especially Facebook) often contained AI-generated descriptions of the media, which aided our LLM-based classifiers.

To prevent latency differences between the control and treatment arms, all arms including control were configured to return results after 500ms. This latency was a hard requirement for our ranking algorithms, which they met more than 95\% of the time. However total user lag was greater than 500ms because it included client-side lag to fetch the initial batch of posts to send to our ranking server, and was on the order of 1-2 seconds for most platforms (including in the control group, where our server just returned the identity ranking). To prevent user attrition after installation due to added latency, all users were slowly ramped up linearly from zero lag to 1200ms lag over the first two weeks after sign up before re-ranking began (which also ensured that we collected baseline pre-intervention data for all users even if they signed up after the nominal intervention start). 

For each ranking algorithm we graph the number of posts added, number of posts deleted, and normalized ranking change per slate. The normalized ranking change is the absolute change in ranking position for a post, divided by the length of the slate (as given in table \ref{tab:slate_length}). Thus, a value of 0.3 means that posts changed rank by 30\% of the slate length on average, e.g. 15 positions for a Twitter slate of length 50. Added and deleted posts are not counted in this metric. 

All users in each experimental arm started treatment at the same time (except for a minority who enrolled less than 14 days before intervention start, allow at least 14 days of baseline data collection), but with different per-arm dates in September and October. Also, intervention on Facebook started later in October for those arms that added posts. We had intended to start all arms and all platforms at the same time, but were unable to do so due to software development software delays. Each ranker section below lists start dates for each platform.

\subsection{Perspective Arms}

\paragraph{People involved (in authorship order):}
Beth Goldberg, Tin Acosta, Emily Saltz

\paragraph{Theory} This intervention builds on Google Jigsaw’s widely used Perspective API to identify rhetorical attributes associated with constructive, bridging discourse \cite{jigsaw_announcing_2024,schmer-galunder_annotator_2024}. The goal is to promote bridging interactions by reordering posts and comments based on predicted conversational attributes, without adding or removing content.

\paragraph{Implementation}
Two rankers were implemented. One ranker uses an average of positive (bridging) attributes provided by the Perspective API. A second ranker uses this same average minus the average of negative Perspective API attributes (i.e., insult and identity attack) and additional proprietary classifiers that detect rhetorical attributes such as moral outrage and alienation. These attributes are summarized in table \ref{tab:perspective_attributes}. Both rankers reorder existing posts and comments and do not add or remove content. 

The \emph{{\perspectivearm}} ranker ordered items by the average of the positive (i.e. bridging) attributes according to the weights in table \ref{tab:perspective_uprank_weights}. The \emph{{\perspectivearmdown}} ranker ordered items by the average of positive attributes minus the negative attributes, according to the weights in table \ref{tab:perspective_uprank_downrank_weights}. The weights were chosen according to the formula $bridging - \frac{1}{2} persuasion - \frac{1}{2} toxicity$  The \textit{Scapegoating}, \textit{Moral outrage} and \textit{Alienation} attributes were further averaged together because these three classifier outputs were strongly correlated.

\paragraph{Classifiers used:} 
The ranker uses classifiers for the experimental attributes of the Google Jigsaw Perspective API. An encoder-only version of Google’s PaLM~2 model was fine-tuned using human ratings of comments from online forums including Reddit, Wikipedia, The New York Times, and the Civil Comments dataset (N = 11{,}973). The latter annotations of the Civil Comments dataset were published in a 2024 benchmark dataset \cite{schmer-galunder_annotator_2024}.

\paragraph{Posts Added} 
These rankers did not add any new posts.

\paragraph{Posts replaced/deleted?} 
None.

\paragraph{Posts, comments, both?} Both.

\begin{figure}
    \centering
    \includegraphics[width=\linewidth]{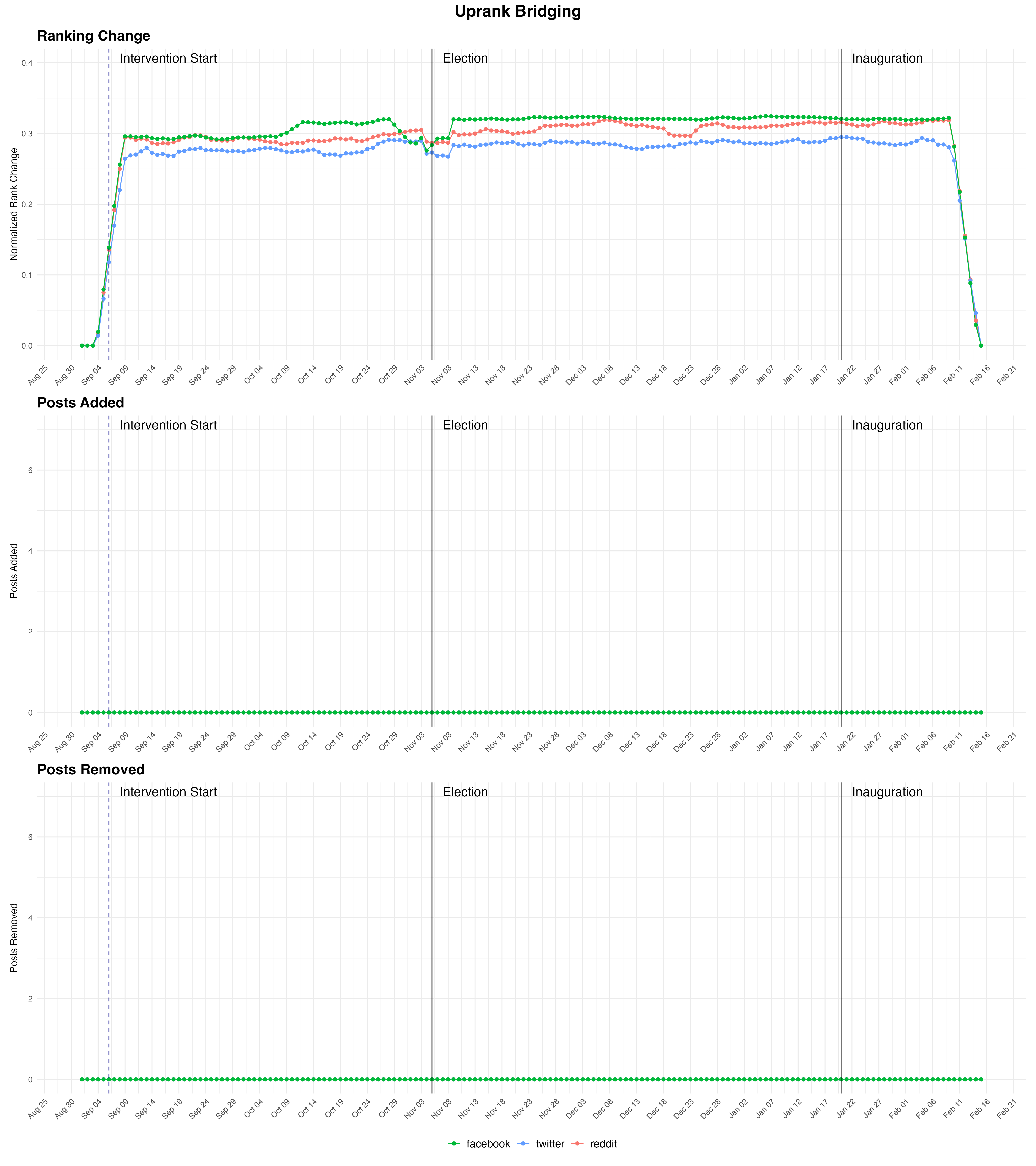}
    \caption{\textbf{Timeline of post rank change, posts added, and posts deleted for \emph{{\perspectivearm}}}, 5 day moving average. Normalized rank change is the average numerical rank change for each item, divided by the length of the item slate.}
    \label{fig:perspective_uprank_timeline}
\end{figure}

\begin{figure}
    \centering
    \includegraphics[width=\linewidth]{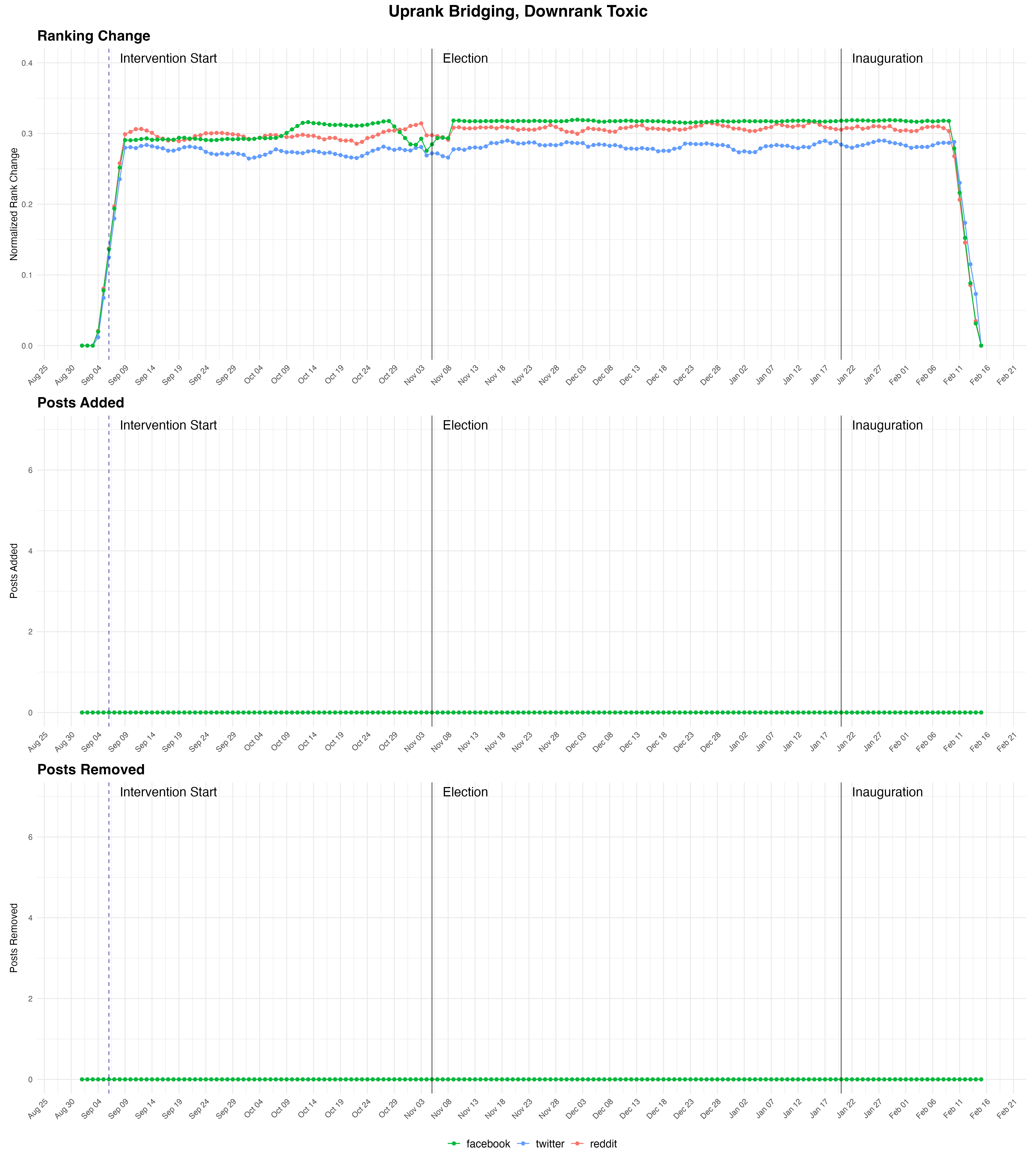}
    \caption{\textbf{Timeline of post rank change, posts added, and posts deleted for \emph{{\perspectivearmdown}}}, 5 day moving average. Normalized rank change is the average numerical rank change for each item, divided by the length of the item slate.}
    \label{fig:perspective_uprank_downrank_timeline}
\end{figure}

\paragraph{Resulting changes} Figures \ref{fig:perspective_uprank_timeline} and \ref{fig:perspective_uprank_downrank_timeline} show the timeline of post rank change, posts added and deleted for \emph{{\perspectivearm}} and \emph{{\perspectivearmdown}}, respectively. As noted above, these rankers do not add or remove content, so the bottom two panels in each figure are zero. The normalized rank change (measured as the average numerical rank change for each item, divided by the length of the item slate) hovers around 0.3 for both rankers, meaning items changed rank by 30\% on average ---a sizable change.  

\paragraph{Where this ranker runs}
All pages with algorithmically ranked posts or comments.

\paragraph{Change log}
Intervention start date: 2024-09-06.\\
Dates when significant changes were made to the algorithm: N/A.\\
Dates where bugs occurred and were/were not fixed: N/A.

\subsection{{\diverseapproval}}

\paragraph{People involved (in authorship order): } 
Luke Thorburn, Soham De, Kayla Duskin, Hongfan Lu, Smitha Milli, Martin Saveski

\paragraph{Theory} 
This ranker approximates the “diverse approval” approach to bridging-based ranking [64] for civic content; namely substituting in civic content that is deemed valuable or engaging by multiple groups who typically disagree. This basic heuristic is used in the Community Notes algorithm on X and other platforms [94], in the civic engagement tool Polis~\cite{small2021polis}, in writing tools that help reach politically diverse audiences~\cite{saveski2022engaging}, and previously by Meta as a signal for comment ranking.\footnote{see https://bridging.systems/facebook-papers/ for the leaked document describing this work} In theory, content that elicits this pattern of engagement is more likely to be written in a way that’s mindful of how it will land with the “other side,” and to represent a form of common experience.

\paragraph{Implementation} 
In each ranking request, all civic posts (defined as posts about political or social issues) were replaced with posts that were both civic and ``bridging,'' defined as posts that would be found “valuable, interesting, and engaging” for both average Republicans and Democrats, according to an LLM rater. In cases where less than 15\% of posts in the request were civic, additional civic and bridging posts were inserted at random positions to attain a minimum dosage of 15\% across treatment arms.

\paragraph{Classifiers} 
The ranker used three classifiers: a civic classifier, a bridging “scorer” (ordinal multi-class classifier), and a national interest classifier.

\begin{itemize}
    \item The civic classifier was a version of PoliBERTweet \cite{kawintiranon2022polibertweet} fine-tuned using a set of 31,359 human-labeled X posts from \cite{milli2025engagement}. We added a randomly initialized binary classification head which was a single dense layer with layer norm and dropout on top of the RoBERTa backbone.
    \item The bridging scorer was implemented by prompting OpenAI’s GPT-4o-mini to score the text of each post on a scale of 1 to 5 reflecting the degree to which an average Republic or Democratic would find the post ``valuable, interesting, and engaging''. A post is considered ``bridging'' if it scored at least 3 from both groups. Posts with higher bridging scores--defined as the lower of the two party-specific scores--were preferentially sampled for insertion.
    \item To focus on nationally relevant content, from 14 Nov onwards we also filtered out local politics (city or state level issues) by prompting GPT-4o-mini with: ``Would the content of the following social media post be of national interest to a U.S. audience?'' 
\end{itemize}
\paragraph{LLM classifier prompt and validation} 
The civic classifier achieved 89.3\% accuracy (F1 score=.73) on a 15\% held-out set of the human-labeled X posts from \cite{milli2025engagement}. The bridging scorer and the national interest classifier were not systematically validated.

\paragraph{Posts added} 
The inventory of posts eligible for insertion was created by regularly collecting posts from a predefined set of sources on each platform, composed broadly of low-slant news media, civil society organisations, state governors and national legislators, as well as the same list of sources used in the {\surprising} ranker. To be eligible for insertion in a given ranking response, a post from this inventory needed to be civic, bridging, recent, and not previously seen by the user. Additionally, we inserted at most one item per source, and removed duplicate posts in each ranking request.

\paragraph{Number of posts added} 
Variable. Additional posts were inserted as needed to ensure that at least 15\% of posts in a ranking request were civic.

\paragraph{Posts replaced/deleted} 
Civic posts were replaced with civic and bridging posts, and additional posts were inserted when necessary to reach the minimum dosage of 15\% civic content.

\paragraph{Resulting changes}

\begin{figure}
    \centering
    \includegraphics[width=\linewidth]{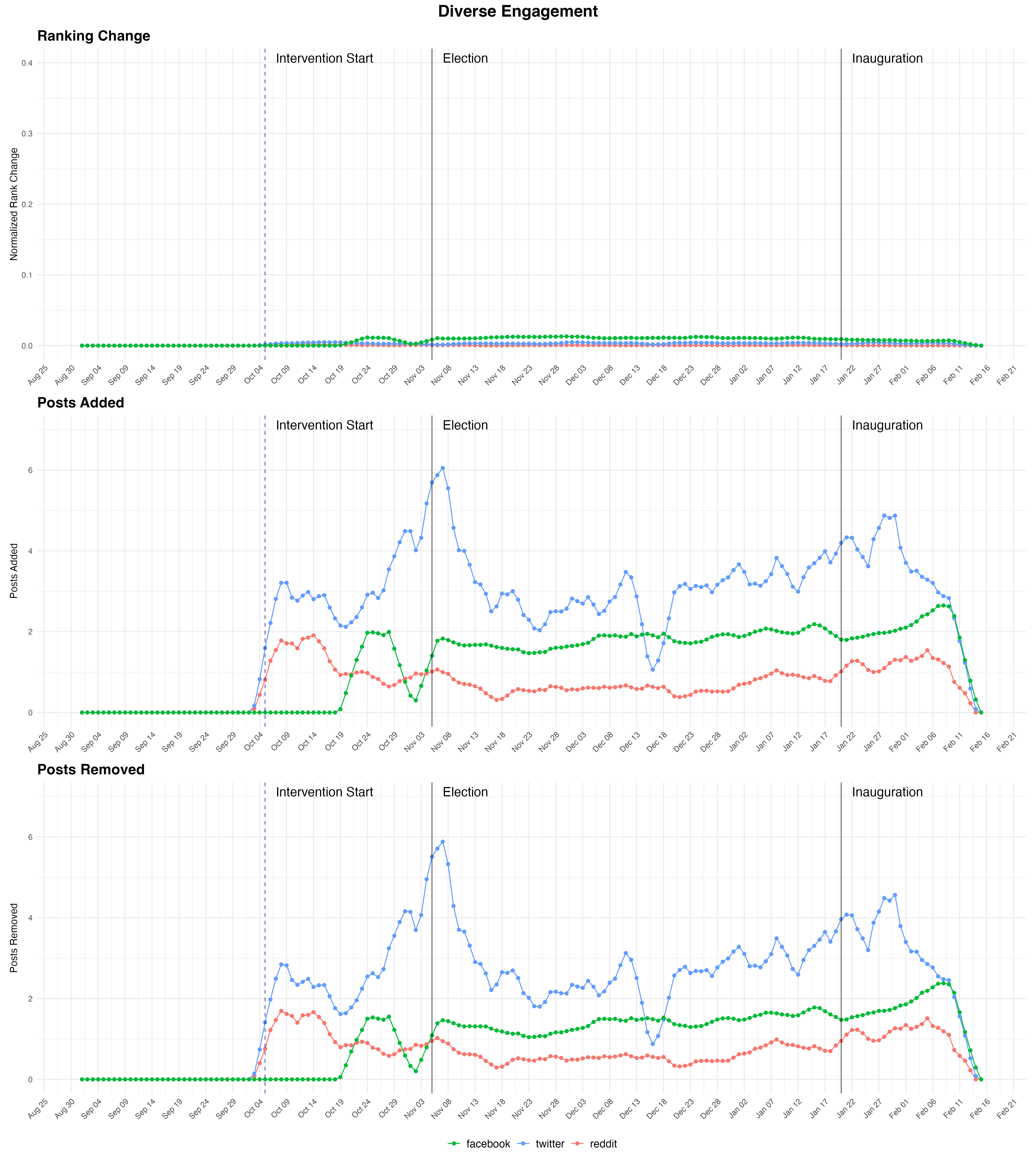}
    \caption{\textbf{Timeline of post rank change, posts added, and posts deleted for \emph{{\diverseapproval}}}, 5 day moving average. Normalized rank change is the average numerical rank change for each item, divided by the length of the item slate.}
    \label{fig:diverseapproval_timeline}
\end{figure}

Figure \ref{fig:diverseapproval_timeline} shows the rank change, added and removed posts over the course of the experiment, 5 day moving average. Normalized rank change is the average numerical rank change for each item, divided by the length of the item slate. Compared to \emph{{\perspectivearm}} and \emph{{\perspectivearmdown}}, this intervention leads to minimal changes in normalized rank change because because \emph{{\diverseapproval}} mostly operates by replacing content, which does not shift the rank of the remaining items. This is reflected in the number of added and removed items mostly mirroring each other. The exception to this is timelines that do not already include at least 15\% civic content. This plot shows that user timelines usually already have at least this fraction of civic content, because there is minimal change in the normalized rank and only small differences between posts added and removed at any point in time.

\paragraph{Personalization} 
None. Added posts were selected from a global pool shared between all users.

\paragraph{Where in the feed are posts inserted?} 
Civic posts were replaced at their original positions. If less than 15\% of the posts in the request were civic, additional posts were inserted at random positions to attain a ``minimum dosage'' of 15\%.

\paragraph{Where this ranker runs}
The ranker intervened only on the main feeds for each platform: the X engagement-based ``For you'' feed, the Facebook feed, and the Reddit homepage. It did not intervene on comment threads or topic-specific subreddits..

\paragraph{Posts, comments, both? } Posts only.

\paragraph{Change log:} Intervention start dates: 2024-10-05 (X/Twitter, Reddit) and 2024-10-21 (Facebook).\\
Dates when significant changes were made to the algorithm: \\
\begin{itemize}
    \item 12 Oct, 12pm PST: Increased the minimum \% of civic posts in ranking responses from 15\% to 20\%.
    \item 21 Oct, 1pm PST: Reduced min \% of civic posts back from 20\% to 15\%.
    \item 26 Oct, 9am PST: Expanded the set of news sources in our inventory to include those with a wider range of political slants (as measured by Media Bias Fact Check ratings).
    \item 7 Nov, 5pm PST: Reduced the maximum age of inserted posts from 5 days to 1 day, and enforced that they be posted after Nov 6 (to avoid posts from before the election outcome seeming particularly out of date).
    \item 14 Nov, 1pm PST: Added the national interest classifier.
    \end{itemize}
Dates where bugs occurred and were/were not fixed: \\
21 Oct, 1pm PST: Fixed bug where we only intervened on x.com (not twitter.com).

\subsection{{\surprising}}

\paragraph{People involved (in authorship order)} 
Siddarth Srinivasan, Max Spohn, Peter Mason, Gauri Jain, Yanchen Jiang, Sara Fish, Alexis Tatore, Safwan Hossain, Manuel Wüthrich

\paragraph{Theory} 
This algorithm recommends content that challenges stereotypes of political in- and out-group members while meeting basic quality thresholds. A growing literature in social psychology and political science has identified misperceptions as an important driver of partisan animosity - and a promising lever to reduce polarization (e.g. \cite{voelkel2024megastudy}). The algorithm is designed to address three types of perceptions:
\begin{itemize}
    \item People perceive out-group members to hold more extreme \cite{levendusky_misperceptions_2016,robinson1995actual} and less varied \cite{dias2025american,zimmerman2022political} political opinions than they really do.
    
    \item People also hold incorrect second-order beliefs, expecting out-group members to view the in-group as more extreme, less human, and less democratic \cite{lees2020inaccurate,landry2023reducing,braley2023voters}.
    
    \item In a case of pluralistic ignorance, people perceive in-group members to hold social norms that punish engagement with opposing partisans and viewpoints, despite the well-documented reputational benefits \cite{heltzel2021seek,yeomans2020conversational}.
\end{itemize}

We inject posts that challenge political stereotypes to shift these perceptions. We find that many such posts come from “surprising validators”  \cite{glaeser_why_2013} - in-group members with high source credibility that challenge in-group beliefs \cite{berinsky2017rumors} and out-group members who contribute to the in-party dialogue \cite{ozer_partisan_2022}. By shifting perceptions of out-group members’ extremity and thoughtfulness, as well as in-group members' willingness to engage with out-group viewpoints, we attempt to reduce affective polarization.

\paragraph{Implementation:} 
The algorithm fetches posts from a pre-selected list of sources on each platform, uses GPT-4o to rate posts along multiple dimensions of ideology and quality, filters them for quality, scores and sorts the post by how much they challenge political stereotypes, and queues them to be recommended. When the user requests a new feed, we identify the top 20 percent most political posts using a fast RoBERTA-based political classifier, then replace them with these pre-selected posts. Inserted posts alternate between liberal- and conservative-leaning to ensure a balanced feed.

\paragraph{Classifiers} 
{\surprising} uses two different classifiers. At serving time, a fast RoBERTa-based classifier from \cite{askari2024incentivizing} is used to decide which existing posts in the user's feed are political, and therefore eligible for replacement. Offline, a GPT-4o based classifier with a single long prompt classifies scraped posts along a variety of ``quality'' dimensions as well as whether it challenges perceptions and political stereotypes. Each scraped post receives probability scores (ranging from 0 to 1) across several content dimensions, and posts are eligible for addition if they fall within acceptable thresholds. Thresholds were set by team members hand labeling whether posts with different scores were suitable for adding to feeds. The filtering dimensions are:

\textbf{Universal Filters (applied across all platforms)}
\begin{itemize}
    \item  NSFW content (threshold $<$ 0.1): Posts flagged as containing adult or inappropriate material were excluded using a strict threshold to ensure workplace-appropriate content.
    \item Niche content (threshold $<$ 0.4): Posts addressing highly specialized topics unlikely to be of broad interest were filtered out.
    \item Frivolous content (threshold $<$ 0.4): Posts lacking substantive informational value were removed.
    \item Outrage-bait (threshold $<$ 0.66): Posts designed primarily to provoke emotional reactions were deprioritized.
    \item Conspiratorial content (threshold $<$ 0.7): Conspiratorial posts were filtered.
    \item Pessimistic framing (threshold $<$ 0.66): Excessively negative or doom-oriented content was limited.
    \item Political orientation (retained if score $>$ 2.1 and $<$ 5.9): Posts with strongly partisan framing at either end of the political spectrum were excluded.
\end{itemize}

\textbf{Platform-Specific Filters}
\begin{itemize}
    \item Reddit: Satirical content (threshold $<$ 0.75) was filtered, as satire on this platform often requires community-specific context.
    \item Twitter/X: Posts required sufficient standalone context (threshold $>$ 0.75) and were filtered for sarcasm (threshold $<$ 0.75), given the platform's brevity constraints and conversational tone.
    \item Facebook: Similar context and sarcasm requirements as Twitter, with an additional filter removing logistical posts (threshold $<$ 0.2) such as event announcements or scheduling content. 
\end{itemize}

\textbf{Final post scoring}
The same prompt also scores scraped posts on two other dimensions, combined to create a single final score.
\begin{itemize}
    \item A score representing how much the post challenges political stereotypes, in 0-1. We prompt the model to flag ``ideologically surprising" content with a few positive and negative examples of real headlines/posts gathered from an initial batch of content.
    \item A source-post mismatch'' score, in the range 0-0.9. We rate the post’s general political orientation and position on the specific issues in the baseline survey (if the post is rated relevant to that issue). We then compute a 0-1 ``post similarity score" as the similarity between the post’s ideological profile and the user’s ideological profile (based on their survey responses). Next, we compute a 0-1 ``source similarity score" as the average post similarity score across posts from a given source (subreddit, twitter account, Facebook page). We select content with high source similarity scores and moderate post similarity scores (the user agrees with the source but not the post), as well as content with high post similarity scores and moderate source similarity scores (the user agrees with the post but not the source). 
\end{itemize}
The post's final score is the maximum of these two scores. Posts are queued for insertion ordered by this final score. Because the maximum source-post mismatch score is approximately 0.9, challenging stereotypes posts are served first.

\paragraph{GPT-4o prompt}

{\footnotesize
\begin{verbatim}
You will be given a post from Facebook. Your task is to explain the post along
with any associated image(s), as well as provide some additional analysis as
specified below. The goal is to identify what kinds of individuals are most
aligned with the content and views expressed in the post. If the post lacks
political/cultural/civic content (broadly construed), you may acknowledge this
in the relevant fields. Provide a JSON response in exactly the following format,
and do not add any other fields.

{
  "explanation": <explain the post and any attached images; watch out for
  sarcasm/ironic posts>,

  "tone": <analyze the tone and sentiment of the post>,

  "ideological_analysis": <analyze the post for its political/ideological
  content, and note any groups/ideological views being promoted or criticized
  (explicitly or implicitly). Even if a post is facially neutral, be sure to
  note if it is likely to be viewed especially favorably or skeptically by any
  particular political identity/group given the broader context of U.S.
  political and cultural opinions. Alternatively, note if the post would have
  broad buy-in from most Americans, except for the strongest partisans/most
  polarized.>,

  "has_enough_context": <true if the point of the post can be understood by a
  general U.S. audience, false otherwise>,

  "is_political_social": <true if the post is political in nature (broadly
  construed, includes economic/social/culture war issues), false otherwise>,

  "is_ideologically_surprising": <Scoring guideline: Determine if the post
  contains ideologically surprising information that challenges ideological
  stereotypes or pre-existing political/cultural beliefs. true if the post
  contains ideologically surprising information. Examples:
    - A traditionally right-leaning individual making "pro-choice" statements.
    - A Democratic politician rejecting reparations for African-Americans.
    - "MAGA Patriots" actively opposing white supremacists at a Trump rally.
    - A well-known liberal figure expressing skepticism about immigration.
    - Texas producing more solar energy than California, contrary to
      expectations.
    - A post showing strong support for both LGBT rights and the U.S. military.
    - A prominent conservative expressing support for a Democratic politician.
    false if not ideologically surprising or if it reinforces typical
    ideological views. Examples:
    - A Republican politician advocating for lower taxes.
    - A liberal-leaning individual advocating for stricter gun control measures.
    - Environmental activists burning down a fossil fuel office.
    - Non-ideological content, such as a woman reuniting with her long-lost
      twin sisters.>,

  "political_orientation": int{1-7} or null <
    7 - Extremely liberal.
    6 - Liberal.
    5 - Slightly liberal.
    4 - Moderate.
    3 - Slightly conservative.
    2 - Conservative.
    1 - Extremely conservative.
    null - Not applicable.>,

  "opinion_trump": int{1-5} or null <
    5 - Highly favorable views of Donald Trump.
    4 - Somewhat favorable views of Donald Trump.
    3 - Neutral views towards Donald Trump.
    2 - Somewhat unfavorable views of Donald Trump.
    1 - Strongly unfavorable views of Donald Trump.
    null - Not applicable.>,

  "opinion_media": int{1-5} or null <
    5 - Strongly trusts mainstream media.
    4 - Somewhat trusts mainstream media.
    3 - Neither trusts nor distrusts mainstream media.
    2 - Somewhat distrusts mainstream media.
    1 - Strongly distrusts mainstream media.
    null - Not applicable.>,

  "opinion_immigration": int{1-5} or null <
    5 - Favors greatly increasing immigration.
    4 - Favors somewhat increasing immigration.
    3 - Favors keeping immigration levels about the same.
    2 - Favors somewhat decreasing immigration.
    1 - Favors greatly decreasing immigration.
    null - Not applicable.>,

  "opinion_abortion": int{1-5} or null <
    5 - Strongly supports a legal right to abortion.
    4 - Somewhat supports a legal right to abortion.
    3 - Neither supports nor opposes.
    2 - Somewhat opposes a legal right to abortion.
    1 - Strongly opposes a legal right to abortion.
    null - Not applicable.>,

  "opinion_israel": int{1-5} or null <
    5 - Strongly supports Israel.
    4 - Somewhat supports Israel.
    3 - Neither supports nor opposes Israel.
    2 - Somewhat opposes Israel.
    1 - Strongly opposes Israel.
    null - Not applicable.>,

  "opinion_climate": int{1-5} or null <
    5 - Extremely concerned about climate change.
    4 - Somewhat concerned about climate change.
    3 - Moderately concerned about climate change.
    2 - Only a little concerned about climate change.
    1 - Not at all concerned about climate change.
    null - Not applicable.>,

  "opinion_racism": int{1-5} or null <
    5 - Racism against Black Americans is a major problem.
    4 - Racism against Black Americans is a significant problem.
    3 - Racism against Black Americans is a moderate problem.
    2 - Racism against Black Americans is only a slight problem.
    1 - Racism against Black Americans is not a problem at all.
    null - Not applicable.>,

  "opinion_military": int{1-5} or null <
    5 - Strongly supports the U.S. military.
    4 - Somewhat supports the U.S. military.
    3 - Neutral towards the U.S. military.
    2 - Somewhat opposes the U.S. military.
    1 - Strongly opposes the U.S. military.
    null - Not applicable.>,

  "opinion_economy": int{1-5} or null <
    5 - Feels extremely positive about the current U.S. economic situation.
    4 - Feels somewhat positive.
    3 - Feels neither positive nor negative.
    2 - Feels somewhat negative.
    1 - Feels extremely negative.
    null - Not applicable.>,

  "is_niche": <true if the post is very niche, lacks enough context, too
    technical, or is not relevant to a general U.S. audience, false otherwise>,
  "is_logistical": <true if the post primarily serves to advertise an event or
    other media without much commentary, false otherwise>,
  "is_frivolous": <true if the post is frivolous, trivial, or lacks substance,
    false otherwise>,
  "is_sarcastic": <true if the post is sarcastic or satirical, false
    otherwise>,
  "is_outrage": <true if the post is inflammatory, antagonistic, or provokes
    outrage, resentment, or strong negative emotion, false otherwise>,
  "is_conspiratorial": <true if the post is sincerely conspiratorial or
    promotes/reflects a lack of social trust, false otherwise>,
  "is_pessimistic": <true if the post reflects or promotes a negative or
    pessimistic outlook, false otherwise>,
  "is_bridging": <true if the post promotes trust, empathy, or understanding
    across the political/ideological spectrum in particular, false otherwise>,
  "is_nsfw": <true if the post contains NSFW content, false otherwise>
}

Here is the Facebook post:

{post}
\end{verbatim}
}

\paragraph{Posts added} 
Posts to be added are fetched from a pre-selected list of sources on each platform at regular intervals, then annotated using GPT-4o as described above. Only those posts that pass the quality filters and are identified as challenging political stereotypes and/or satisfy source–post mismatch criteria are selected for recommendation. 

For each of the three platforms, we asked ChatGPT and Claude to generate a list of sources (subreddits, Twitter accounts, Facebook pages) from various domains (politics, entertainment, technology, journalism, academia). On Twitter, we additionally crawled the ``following" list of suggested accounts and sampled accounts with large followings. We then conducted an initial screen to discard sources with low “yield”, i.e., sources whose content was unlikely to be recommended by our algorithm (due to low quality or low frequency of content that challenged political stereotypes). This left us with 139 Facebook pages and groups, 413 Twitter sources, and 95 Subreddits.

The following sources yielded the most top-scoring posts under our algorithm in absolute terms. Note that this is not necessarily representative of the average score of posts from these sources.

\begin{itemize}
    \item On Reddit: r/centrist, r/vaushv, r/conservative, r/neoliberal, r/uspolitics
    \item On Twitter: Mediaite, The Hill, Fox News, DC Examiner, and Politico.
    \item On Facebook: The Hill, Fox News, The Daily Beast, The Daily Caller, Huffington Post Politics
\end{itemize}

\paragraph{Number of posts added} 
Approximately 20\% of feed items.

\paragraph{Posts replaced/deleted} 
On each request, the 20\% most political posts in the user’s feed are replaced with queued posts. 

\paragraph{Resulting changes}
Overall, there is no change in normalized rank change since this ranker only replaces items, leaving the rank of existing items unchanged. 

\paragraph{Personalization} 
This algorithm uses two sources of posts. Posts are selected if they challenge political stereotypes, and these are not personalized. Posts are also selected in a personalized way, by comparing the user's political position on nine issues (known from their baseline survey answers) to the political position of posts and sources (identified by the GPT-4o classifier). This is used to select content where the user agrees with the source but not the post, or where the user agrees with the post but not the source.

\paragraph{Where this ranker runs}
The ranker intervened only on the X engagement-based timeline, the Facebook feed, and the Reddit homepage. It did not intervene on comment threads or topic-specific subreddits.

\paragraph{Posts, comments, both? } Posts only.

\paragraph{Change log:}
Intervention start dates: 2024-10-06 (Twitter/X, Reddit) and 2024-10-21 (Facebook).\\
Dates when significant changes were made to the algorithm: N/A.\\
Dates where bugs occurred and were/were not fixed: Worker that managed annotation of scraped posts with GPT-4o failed starting around Jan 12 and was fixed on Jan 28. During this time we served queued posts that were more than three days old, which would otherwise have expired.

\subsection{{\addnews}}

\paragraph{People involved (in authorship order):} Muhammed Haroon, Anshuman Chhabra, Jonathan Stray, Magdalena E. Wojcieszak

\paragraph{Theory} 
This intervention adds posts from verified and ideologically balanced news outlets to increase users’ exposure to factual public affairs information toward increasing knowledge, efficacy, political interest, and belief accuracy.

\paragraph{Implementation}
The algorithm adds posts from verified and ideologically balanced news outlets selected by overlapping high-reliability ranking lists from NewsGuard and Ad Fontes. Posts are selected using a weighted sampling strategy across ideological categories and are personalized using users’ prior histories. User embeddings are generated in the background, candidate posts are stored in a Redis database, and real-time requests are served from this database. Crawling and computation occur in background processes at regular intervals of three hours.

\paragraph{Classifiers} Sentence-transformer models are used to generate embeddings for personalization. The \verb|stella_en_400M_v5| model from the SentenceTransformers library is used for efficiency and low resource consumption. 

\paragraph{Posts added} 
We selected 95 verified and ideologically balanced news outlets by overlapping high-reliability sources listed by NewsGuard and Ad Fontes.
For Twitter/X and Facebook, posts are scraped from the accounts operated by these sources. For Reddit, posts are scraped from news subreddits, and only posts whose external URLs are from the selected news sources are used. Posts are selected using a weighted sampling strategy that assigns 25\% weight to left-leaning sources, 50\% weight to moderate sources, and 25\% weight to right-leaning sources.

All scraping was done through the Apify service. Only scraped posts which are within 48 hours of the newest scraped post are eligible for addition to the user's feed, and a 48-hour shown-posts cache prevents the same item from being recommended multiple times within that window. These two rules allow graceful degradation of the user experience if the scraper stops running, which happened on Facebook around the week of October 29.

\paragraph{Number of posts added} 
An upper bound is imposed such that no more than 15\% of a user’s daily posts are injected by the algorithm.

\paragraph{Posts replaced/deleted} 
None.

\begin{figure}
    \centering
    \includegraphics[width=\linewidth]{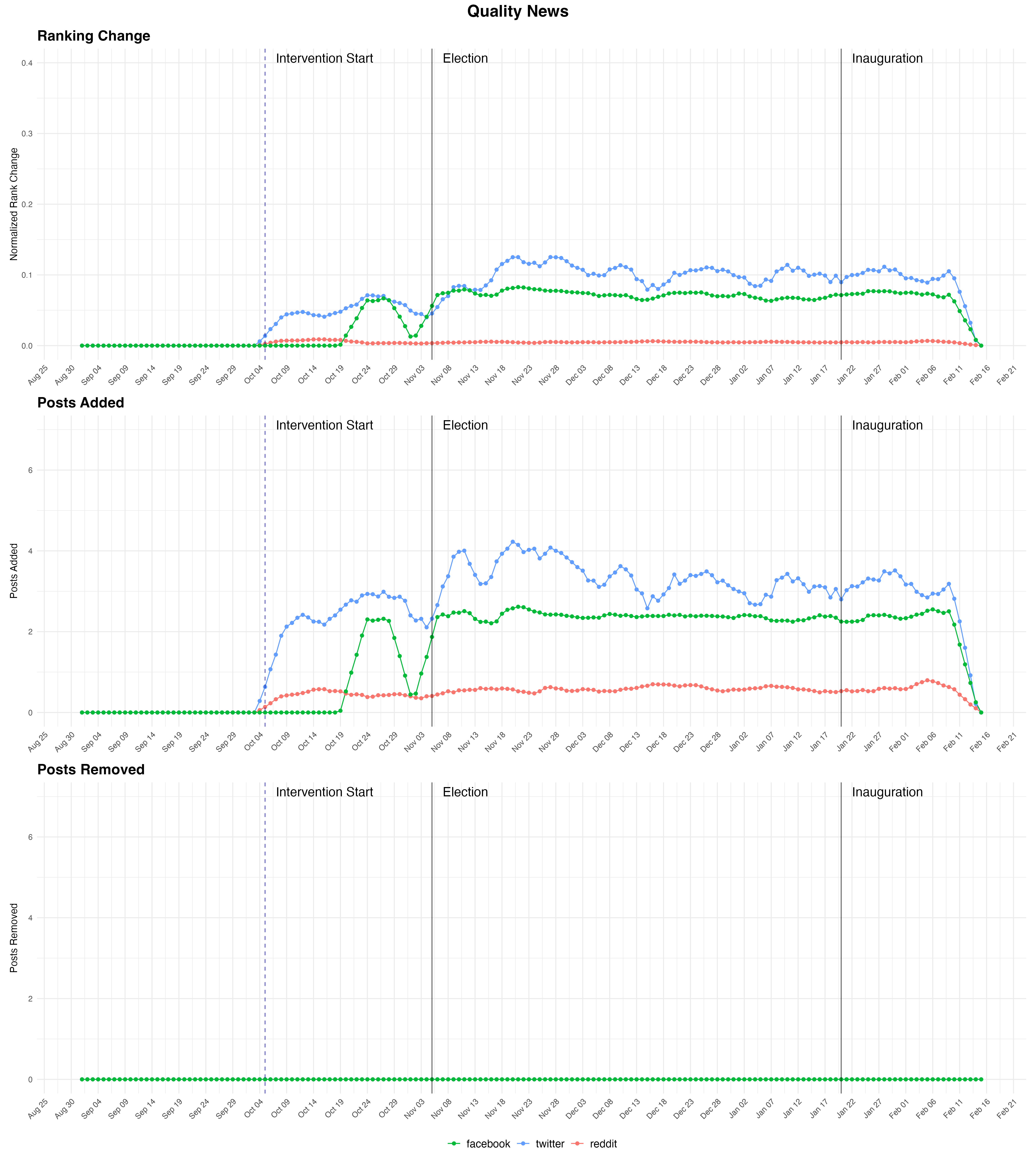}
    \caption{\textbf{Timeline of post rank change, posts added, and posts deleted for \emph{{\addnews}}}, 5 day moving average.}
    \label{fig:addnews_timeline}
\end{figure}

\paragraph{Resulting Changes}
Figure \ref{fig:addnews_timeline} shows the patterns of rank changes, added and removed content for this ranker.
Since \emph{{\addnews}} does not replace/remove any content, the third panel is flat.

\paragraph{Personalization} 
Personalization is achieved by identifying scraped news posts whose content is similar to the participant’s prior viewing history. In the offline worker, for each user we retrieve the posts served by the platform from each of the last 100 ranking requests. We then sample 50 of these posts randomly to create a reference set (filtering out any post shorter than 20 characters). Each sampled served post and each candidate scraped post is embedded using the \verb|stella_en_400M_v5| sentence transformer model. For each sampled post we find the five most similar candidate  posts by directly computing all cosine similarities. We score candidate posts by counting the number of sampled posts for which they are one the top five closest posts. The top 50 highest scoring posts are randomly shuffled and queued for addition to the user's feed. We do not try to preserve ideological diversity in this personalization step, relying on the diversity of scraped sources for balance. This whole process runs several times a day for each platform for each user. 

\paragraph{Where this ranker runs}The ranker intervened only on the X engagement-based timeline, the Facebook feed, and the Reddit homepage. It did not intervene on comment threads or topic-specific subreddits.

\paragraph{Posts, comments, both? } Posts only.

\paragraph{Change log:} Intervention start dates: 2024-10-05 (Twitter/X, Reddit) and 2024-10-21 (Facebook).\\
Dates when significant changes were made to the algorithm: N/A.\\
Dates where bugs occurred and were/were not fixed: N/A.

\section{Participant Recruitment and Sample}\label{app:Recruitment}

Because we needed to recruit many people quickly and screen in rates were low, participants were mostly recruited by presenting a screening survey to as many people as possible using more than a dozen U.S. consumer market research firms. A significant number came from a research panel maintained by Forthright, and a smaller number from self-serve research platforms CloudResearch and Positly. A small number came from social media ads, before we determined that this was not a cost effective recruiting strategy. The participant source breakdown is shown in table \ref{tab:recruitment_sources}. The final sample of 9,386 are those of the 9,701 who also completed the baseline survey.

Participants were recruited through a multi-stage process. First, they had to pass the screening survey shown in table \ref{tab:screener}  which asked whether they were desktop users of Facebook, Reddit and/or X/Twitter on the Chrome browser. Participants were paid for this screening survey even if they were not eligible. We did not have control over the amount of this payment as it varied from vendor to vendor based on our negotiated price and their pass-through but it was small, typically less than \$1 US. Those who were eligible for the study were then redirected to socialmedialab.study and asked to install the extension. Those who installed the extension were asked to complete the baseline survey. If they installed the extension but didn't complete the baseline survey, we sent occasional email and browser notification reminders. The overall conversion rate from completing the screening survey to inclusion in the sample study (defined as completion of baseline survey and the extension kept installed for at least one week) was 3.7\%, meaning that we contacted over 250,000 people to obtain our final sample. Our total recruitment cost was over \$400,000, with the majority ($>80$\%) going toward user acquisition rather than payment for surveys. 

We learned some important lessons from this massive recruitment effort. First, we needed so many people to sign up so fast that we essentially saturated the consumer market research industry for a few weeks, meaning that there is probably no easy way to recruit participants faster even if money is not an issue. Second, we probably could have benefited greatly from better participant communication. Although we stated in multiple places during the recruitment pipeline that people would receive \$5 for each of two follow up surveys months later, many people were still confused about when they would get paid, and often emailed asking. This confusion may be why many people uninstalled the extension immediately after installation. Were we to attempt such a long-term study again, we would try to be even clearer about the payment structure, which may reduce short term attrition. We would also send monthly update emails confirming enrollment and telling people when to expect the next survey and payment, which may reduce long term attrition.

\section{Identification Strategy}\label{app:Identification}

\subsection{Behavioral Outcomes (Daily Panel)}
\paragraph{Pooled treatment arms.}
We estimate the following two-way fixed effects (TWFE) regression:

\begin{equation}
    Y_{i,t} = \alpha_i + \delta_{t, C_i, d} + \beta D_{i,t} + \varepsilon_{i,t},
\end{equation}

where $Y_{i,t}$ is the outcome for user $i$ in period $t$ (measured relative to the intervention start date), $\alpha_i$ represents individual fixed effects, $D_{i,t}$ is an indicator variable equal to 1 for periods on or after the day when the user had treatment switched on. To address the staggered treatment roll-out, as well as the fact that the earliest treatment start was different for different treatment arms, $\delta_{t, C_i, d}$ captures the interaction of period, cohort, and date fixed effects, i.e., a separate day-level period dummy for each cohort defined by its enrollment date. The triple interaction of fixed effects extends the ``stacked regression'' approach to ensuring robustness to the staggered roll-out of the treatment to a case where different treatments -- see \cite{cengiz2019effect} for an application.

For the control group, we set the intervention start date to the latest of the earliest intervention start date in the sample and 14 days after enrollment, ensuring at least two weeks of baseline period.

\paragraph{Individual arms.}

For individual arms, we subset the sample to the relevant individual treatment arm and the control group and set the hypothetical intervention start date, for the control group, to the intervention start date of the respective cohort, meaning that $\delta_{t, C_i, d}$ effectively becomes $\delta_{t, C_i}$.

\paragraph{Heterogeneity.}

We estimate heterogeneity of treatment effects with respect to baseline and demographics characteristics  using the following specification:

\begin{equation}
    Y_{i,t} = \alpha_i + \delta_{t, C_i, d, H_{i}} + \beta D_{i,t} + \phi\,(D_{it} H_{i}) + \varepsilon_{i,t},
\end{equation}

where the dummy $H_{i}$ encodes a baseline or demographic characteristic.

We exploit the opportunity presented by having a large number of daily time periods in the study, and use \cite{driscoll1998consistent} standard errors, which account for both serial and cross-sectional dependence, to increase power.

\subsection{Survey and In-Feed Outcomes}
We estimate the two-way fixed effects (TWFE) regression with three time periods:

\begin{equation}
    Y_{i,t} = \alpha_i + \delta_{t} + \beta D_{i,t} + \varepsilon_{i,t},
\end{equation}

where $Y_{i,t}$ is the outcome for user $i$ at time $t$, and $D_{i,t}$ is an indicator whether the user $i$ had already been treated by time $t$. For individual treatment arms, we restrict the sample to the control group and the respective treatment arm. Heterogeneity analysis, similarly to the behavioral outcomes, adds the term $\phi\,(D_{it} H_{i})$.

Given the relatively low number of time periods, for these specifications we use clustered standard errors.

\section{Attrition}
\label{app:attrition}

\subsection{Survey Attrition}

Overall, our survey completion rate is 53\%. This figure measures the proportion of users who completed either a midline or endline survey out of all who (1) completed a pre-treatment survey and (2) installed the extension and kept it installed for one week. 

We find no evidence that participants assigned to the treatment group completed surveys at different rates compared to the control group. 
Table \ref{tab:survey_attrition} shows the results of several OLS regressions where completion of either a midline or endline survey is the dependent variable and treatment assignment is the independent variable. 
Model (1) of the table shows the basic, bivariate model. The substantively tiny and statistically insignificant coefficient on the treatment dummy indicates that treated participants were not more or less likely than control participants to complete the surveys. 
Model (2) of the table models the predicted probability of completing a survey as a function of a propensity score that combines available demographic and baseline characteristics (gender, race, education, party affiliation, income, baseline social media usage, and baseline polarization), and interacts this predicted survival score with treatment assignment.
The result suggests that participants who are in general more likely to fill out surveys based on these characteristics are equally as likely to fill out the midline or endline survey when assigned to treatment as compared to control. 
Finally, Model (3) models survey taking as a function of each treatment arm individually. Again, no significant differences are observed between treatment arms.

\section{All Survey Questions}\label{app:Survey_Questions}

\noindent \textbf{party id} \cite{iyengar_affect_2012}: Generally speaking, do you usually think of yourself as a Republican, Democrat, Independent, or what?

[Republican, Democrat, Independent, No preference, Other (please specify)]

\vspace{1em}

\noindent \textbf{party strong} (if Republican or Democrat) \cite{iyengar_affect_2012}: Would you call yourself a strong [party id choice] or not a very strong [party id choice]? 

[Strong, Not very strong]

\vspace{1em}

\noindent \textbf{party leaner} (if NOT Republican or Democrat) \cite{iyengar_affect_2012}: Do you think of yourself as closer to the Republican or Democratic Party? 

[Republican, Democratic]

\vspace{1em}

\noindent \textbf{browser mobile}: For each of the platforms below, how often would you say you use social media from your desktop computer (including laptops), as opposed to your mobile device?   For each platform, please use the slider where 0 means only mobile and 100 only desktop. If you do not use the platform, please check ``Not Applicable."

[slider, 0 to 100]

[Facebook, Reddit, X (Twitter)]

\vspace{1em}

\noindent \textbf{gender}: Are you:

[Male, Female, Non-binary, Prefer not to say]

\vspace{1em}

\noindent \textbf{age} What is your age?

[18-24, 25-34, 35-44, 45-54, 55-64, 65+] 

\vspace{1em}

\noindent \textbf{education} What is the highest level of education you have completed?

[Some high school, Graduated high school, Some college, Associate’s degree, Bachelor’s degree, Graduate degree]

\vspace{1em}

\noindent \textbf{ideology} Here is a seven-point scale on which the political views that people might hold are arranged from extremely liberal—point 1—to extremely conservative—point 7. Where would you place yourself on this scale?

[1. Extremely liberal, 2. Liberal, 3. Slightly liberal, 4. Moderate, middle of the road, 5. Slightly conservative, 6. Conservative, 7. Extremely conservative]

\vspace{1em}

\noindent \textbf{income} What is your annual household income?

[Less than \$20,000, \$20,000 to \$34,999, \$35,000 to \$49,999, \$50,000 to \$74,999, \$75,000 to \$99,999, \$100,000 to \$149,999, \$150,000+] 

\vspace{1em}

\noindent \textbf{ethnicity} Which race or ethnicity best describes you? (Please choose only one.)

[American Indian or Alaskan Native, Asian / Pacific Islander, Black or African American, Hispanic, White / Caucasian, Multiple ethnicity/ Other (please specify)] 

\vspace{1em}

\noindent \textbf{socmed use}: Think of the past two weeks. How much time did you spend on social media, on average, per day?

[0–30 minutes, 30–60 minutes, 60–90 minutes, 90–120 minutes, 2–3 hours, 3–4 hours, More than 4 hours]

\vspace{1em}

\noindent \textbf{desktop share}: Is your desktop shared with others?

[Yes, No]

\vspace{1em}

\noindent \textbf{wellbeing} \cite{topp_who-5_2015}: Over the past 2 weeks, how often have you felt the following?

[All of the time, Most of the time, More than half the time, Some of the time, At no time]

\begin{itemize}
\item I have felt cheerful and in good spirits.
\item I have felt calm and relaxed.
\item I have felt active and vigorous.
\item I woke up feeling fresh and rested.
\item My daily life has been filled with things that interest me.
\end{itemize}

\vspace{1em}

\noindent \textbf{ft out} \cite{iyengar_affect_2012}: Please indicate how you feel toward [out party] voters using the scale below. 100 means that you feel very favorably or warm toward them, 0 means that you feel very unfavorable or cold, and 50 means neutral.

[Feeling thermometer: 0 (Cold) – 100 (Warm)]

\vspace{1em}

\noindent \textbf{ft in} \cite{iyengar_affect_2012}: Please indicate how you feel toward [in party] voters using the scale below. 100 means that you feel very favorably or warm toward them, 0 means that you feel very unfavorable or cold, and 50 means neutral.

[Feeling thermometer: 0 (Cold) – 100 (Warm)]

\vspace{1em}

\noindent \textbf{kalmoe mason 3} \cite{kalmoe_lethal_2019}: How much do you feel it is justified for [out party] to use violence in advancing their political goals these days?

[0 (Not at all justified) – 100 (Extremely justified)]

\vspace{1em}

\noindent \textbf{kalmoe mason 1} \cite{kalmoe_lethal_2019}: When, if ever, is it OK for [in party] to send threatening and intimidating messages to [out party] leaders?

[0 (Never OK) – 100 (Always OK)]

\vspace{1em}

\noindent \textbf{meta spv 1} \cite{mernyk_correcting_2022}: How do you think an average [out party] would respond to the following question? \textit{How much do you feel it is justified for [out party] to use violence in advancing their political goals these days?}

[0 (Not at all justified) – 100 (Extremely justified)]

\vspace{1em}

\noindent \textbf{meta spv 2} \cite{mernyk_correcting_2022}: How do you think an average [out party] would respond to the following question? \textit{When, if ever, is it OK for [out party] to send threatening and intimidating messages to [out party] leaders?}

[0 (Never OK) – 100 (Always OK)]

\vspace{1em}

\noindent \textbf{meta wev past} \cite{mernyk_correcting_2022}: Think about the recent presidential election. If [in party candidate, Trump or Harris] was declared the winner of a close, contested election, how likely do you think [out party] voters would have been to engage in violence?

[0 (Not at all likely) – 100 (Extremely likely)]

\vspace{1em}

\noindent \textbf{intergroup empathy} \cite{sirin_group_2016}: How much do you agree or disagree with the following statements?

[1 = Strongly disagree, 2, 3, 4, 5, 6, 7 = Strongly agree]

\begin{itemize}
\item I find it difficult to see things from [out party]' point of view.
\item I think it is important to understand [out party] by imagining how things look from their perspective.
\end{itemize}

\vspace{1em}

\noindent \textbf{gss general trust}: Generally speaking, would you say that most people can be trusted, or that you can't be too careful in dealing with people?

[0 (Most people can be trusted) – 100 (You can't be too careful)]

\vspace{1em}

\noindent \textbf{sd out} \cite{shanto_origins_2019}: How comfortable are you having friends who are [out party]?

[1 (Not at all comfortable) – 7 (Extremely comfortable)]

\vspace{1em}

\noindent \textbf{sd in} \cite{shanto_origins_2019}: How comfortable are you having friends who are [in party]?

[1 (Not at all comfortable) – 7 (Extremely comfortable)]

\vspace{1em}

\noindent \textbf{learned fb} \cite{fast_unveiling_2023}: In the last two weeks, have you learned something that was useful or helped you understand something important on Facebook?

[Yes, No, Not sure]

\vspace{1em}

\noindent \textbf{learned x} \cite{fast_unveiling_2023}: In the last two weeks, have you learned something that was useful or helped you understand something important on X (Twitter)?

[Yes, No, Not sure]

\vspace{1em}

\noindent \textbf{learned reddit} \cite{fast_unveiling_2023}: In the last two weeks, have you learned something that was useful or helped you understand something important on Reddit?

[Yes, No, Not sure]

\vspace{1em}

\noindent \textbf{bftw fb} \cite{fast_unveiling_2023}: In the last two weeks, have you witnessed or experienced content that you would consider bad for the world on Facebook?

[Yes, No, Not sure]

\vspace{1em}

\noindent \textbf{bftw x} \cite{fast_unveiling_2023}: In the last two weeks, have you witnessed or experienced content that you would consider bad for the world on X (Twitter)?

[Yes, No, Not sure]

\vspace{1em}

\noindent \textbf{bftw reddit} \cite{fast_unveiling_2023}: In the last two weeks, have you witnessed or experienced content that you would consider bad for the world on Reddit?

[Yes, No, Not sure]

\vspace{1em}

\noindent \textbf{meaningful fb} \cite{fast_unveiling_2023}: In the last two weeks, have you experienced a meaningful connection with others on Facebook?

[Yes, No, Not sure]

\vspace{1em}

\noindent \textbf{meaningful x} \cite{fast_unveiling_2023}: In the last two weeks, have you experienced a meaningful connection with others on X (Twitter)?

[Yes, No, Not sure]

\vspace{1em}

\noindent \textbf{meaningful reddit} \cite{fast_unveiling_2023}: In the last two weeks, have you experienced a meaningful connection with others on Reddit?

[Yes, No, Not sure]

\vspace{1em}

\noindent \textbf{negative fb} \cite{fast_unveiling_2023}: In the last two weeks, have you personally witnessed or experienced something that affected you negatively on Facebook?

[Yes, No, Not sure]

\vspace{1em}

\noindent \textbf{negative x} \cite{fast_unveiling_2023}: In the last two weeks, have you personally witnessed or experienced something that affected you negatively on X (Twitter)?

[Yes, No, Not sure]

\vspace{1em}

\noindent \textbf{negative reddit} \cite{fast_unveiling_2023}: In the last two weeks, have you personally witnessed or experienced something that affected you negatively on Reddit?

[Yes, No, Not sure]

\subsection{Political Knowledge Questions}

Each time participants took the main survey, they were asked the to review recent news headlines and evaluate whether or not they are true. We gathered five headlines pulled from the top articles on Google news within the past two weeks which were related to political events. We altered two to three of these headlines each week to create a false version (following to the approach used in \cite{allcott2020welfare}). Question wording was as follows:

Of the following news events, which ones do you think are true events that occurred in the last month, and which ones do you think are false and did not occur?

[True, False, Unsure]\\

We presented the following sets of headlines, refreshed regularly throughout the study:\\

\paragraph{Group 1}
\begin{itemize}
  \item (F) Nikki Haley fails to win any Republican primaries before ending her election bid
  \item (T) U.N. Team finds grounds to support reports of sexual violence in Hamas Attack
  \item (F) Alabama governor vetoes IVF protection bill, limiting the protection for fertility services in the state
  \item (T) Biden directs US military to establish aid port in Gaza
  \item (T) Biden and Trump trade attacks in dueling events in Georgia
\end{itemize}

\paragraph{Group 2}
\begin{itemize}
  \item (F) US House votes against \$95 billion aid package for Ukraine, Israel, and Taiwan
  \item (T) Trump Hush money trial commences
  \item (F) ByteDance agrees to sell TikTok, avoiding a US shutdown
  \item (T) USC cancels main commencement ceremony after protests, arrests
  \item (T) Tornadoes devastate parts of Nebraska and Iowa
\end{itemize}

\paragraph{Group 3}
\begin{itemize}
  \item (F) India's Modi reelected in a landslide, increasing his party's majority
  \item (T) Louisiana governor signs bill to classify abortion pills as controlled substances into law
  \item (T) Taiwan's parliament passes bill pushing pro-China changes
  \item (F) Biden details Gaza truce proposal, Hamas rejects it outright
  \item (T) In Shift, Biden Issues Order Allowing Temporary Border Closure to Migrants
\end{itemize}

\paragraph{Group 4}
\begin{itemize}
  \item (F) The U.N. Security Council rejects U.S. cease-fire plan to end the war in Gaza
  \item (T) Inflation slows in May, with consumer prices up 3.3\% from a year ago
  \item (T) Unanimous Supreme Court preserves access to abortion pill mifepristone
  \item (T) Biden tells congressional Democrats he is ``firmly committed'' to staying in 2024 race
  \item (F) Ohio House rejects bill restricting bathroom use for transgender students
\end{itemize}

\paragraph{Group 5}
\begin{itemize}
  \item (F) Supreme Court upholds EPA's interstate air pollution regulation
  \item (F) Labour Party's Keir Starmer concedes defeat as Prime Minister Rishi Sunak and his Conservative Party head towards another sweeping UK election victory
  \item (T) Assassination attempt on Trump at Pennsylvania rally leaves 2 hurt, 2 dead, including shooter
  \item (T) Biden reportedly to push for supreme court term limits and new ethics code
  \item (F) Speaker Johnson calls on Biden to resign after the president announces he won't seek reelection
\end{itemize}

\paragraph{Group 6}
\begin{itemize}
  \item (T) Venezuelan President Nicolas Maduro claims election victory, refuses to publish results
  \item (F) U.S., Russia, and other countries end prisoner swap negotiations after failing to reach agreement
  \item (T) US economy added far fewer jobs than expected in July
  \item (F) Google is not an illegal monopoly, federal court rules
  \item (T) Harris selects Minnesota Gov.\ Tim Walz to be VP running mate
\end{itemize}

\paragraph{Group 7}
\begin{itemize}
  \item (T) Trump campaign hacked, blames Iran for stealing internal communications
  \item (T) Ilhan Omar, member of the `Squad', wins Minnesota Democratic primary
  \item (F) Minouche Shafik Announces She Will Continue Tenure as Columbia President, Seeks to `Reunify' Campus
  \item (T) `The answer is no': Pro-Palestinian delegates say their request for a speaker at DNC was shut down
  \item (F) RFK Jr.\ suspends campaign; endorses VP Harris
\end{itemize}

\paragraph{Group 8}
\begin{itemize}
  \item (F) Supreme Court Agrees to Restart Biden's Latest Student Debt Relief Plan
  \item (F) Harris vows to ban fracking as president
  \item (T) Harris says she'd appoint a Republican to her Cabinet if elected
  \item (T) Trump Says He'll Vote Against Florida's Abortion Rights Measure After Conservative Backlash
  \item (T) US accuses Russian state media of meddling in 2024 election
\end{itemize}

\paragraph{Group 9}
\begin{itemize}
  \item (F) Putin Declares Support for Trump, Dismisses Harris
  \item (F) Ohio police note `some credible reports' of Haitian immigrants harming pets, are investigating further
  \item (T) Donald Trump Rejects A Rematch, Says There Won't Be Another Debate With Kamala Harris
  \item (T) Trump is safe following shooting at Florida golf course; suspect detained
  \item (T) Netanyahu warns Yemen's Houthis face a ``heavy price'' as missile lands in central Israel
\end{itemize}

\paragraph{Group 10}
\begin{itemize}
  \item (T) Mark Robinson Vows to Stay in N.C.\ Governor's Race After CNN Report
  \item (F) Freddie Owens execution delayed after key witness admits he lied
  \item (T) No government shutdown for now: Congress agrees on temporary funding deal through December
  \item (F) Trump says he would run for president again in 2028 if he loses this time
  \item (T) Ukraine's Zelenskiy says US decisive actions could end Russian war next year
\end{itemize}

\paragraph{Group 11}
\begin{itemize}
  \item (T) Gaetz Paid for Sex, Used Drugs While in Congress, Panel Finds
  \item (F) Luigi Mangione Pleads Guilty to State Murder and Terror Charges
  \item (T) Azerbaijan Airlines plane headed to Russia crashes hundreds of miles off course, dozens dead
  \item (F) Donald Trump wins appeal against E Jean Carroll sexual abuse verdict
  \item (T) Mike Johnson re-elected House speaker after GOP mutiny threat dissolves
\end{itemize}

\section{Additional Robustness Checks}

Table \ref{app:robustness_inclusion} shows our main results when defining the inclusion criterias as keeping the extension installed for two days or two weeks, as opposed to our main specification of one week. Table \ref{app:robustness_twitter} shows our main results without any user who used X/Twitter at all during their two week pre-intervention baseline period.


\clearpage

\begin{figure}
    \centering
    \includegraphics[width=\linewidth]{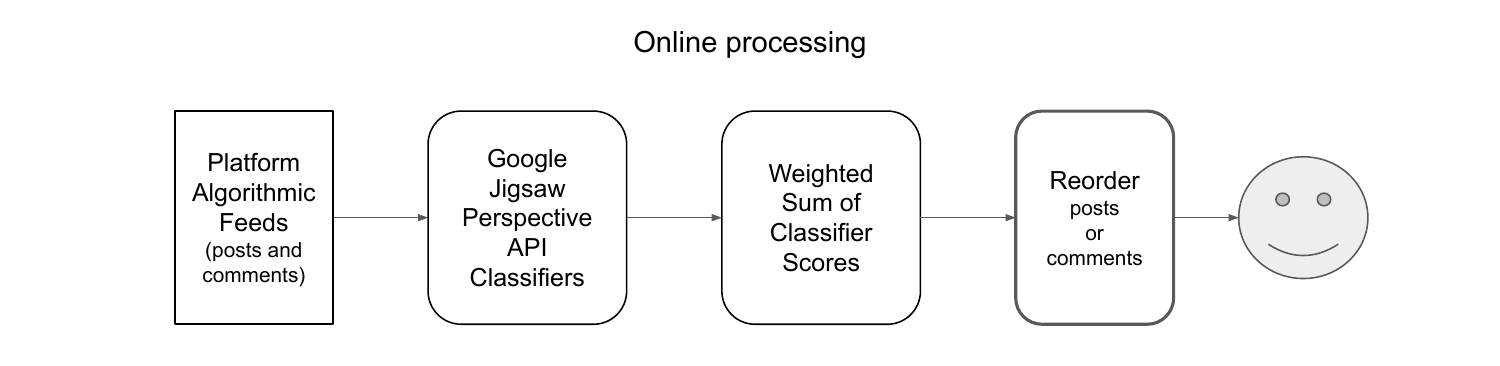}
    \caption{\textbf{Design of the Perspective ranking algorithms.} The \emph{{\perspectivearm}} and \emph{{\perspectivearmdown}} algorithms are identical except for different classifier weights, as listed in tables \ref{tab:perspective_uprank_weights} and \ref{tab:perspective_uprank_downrank_weights}.}
    \label{fig:perspective_design}
\end{figure}

\begin{figure}
    \centering
    \includegraphics[width=\linewidth]{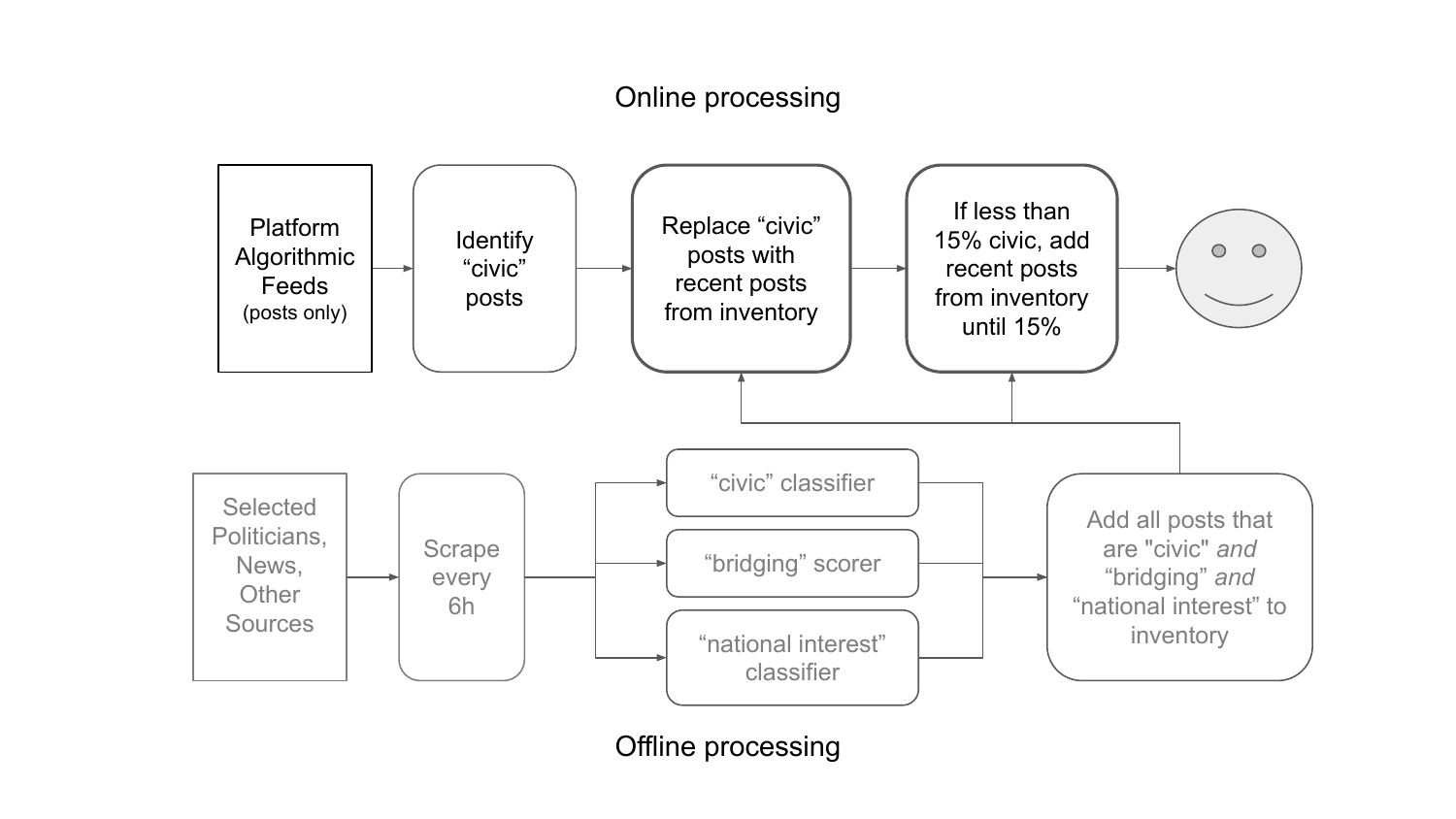}
    \caption{\textbf{Design of the {\diverseapproval} ranking algorithm.}}
    \label{fig:perspective_design}
\end{figure}

\begin{figure}
    \centering
    \includegraphics[width=\linewidth]{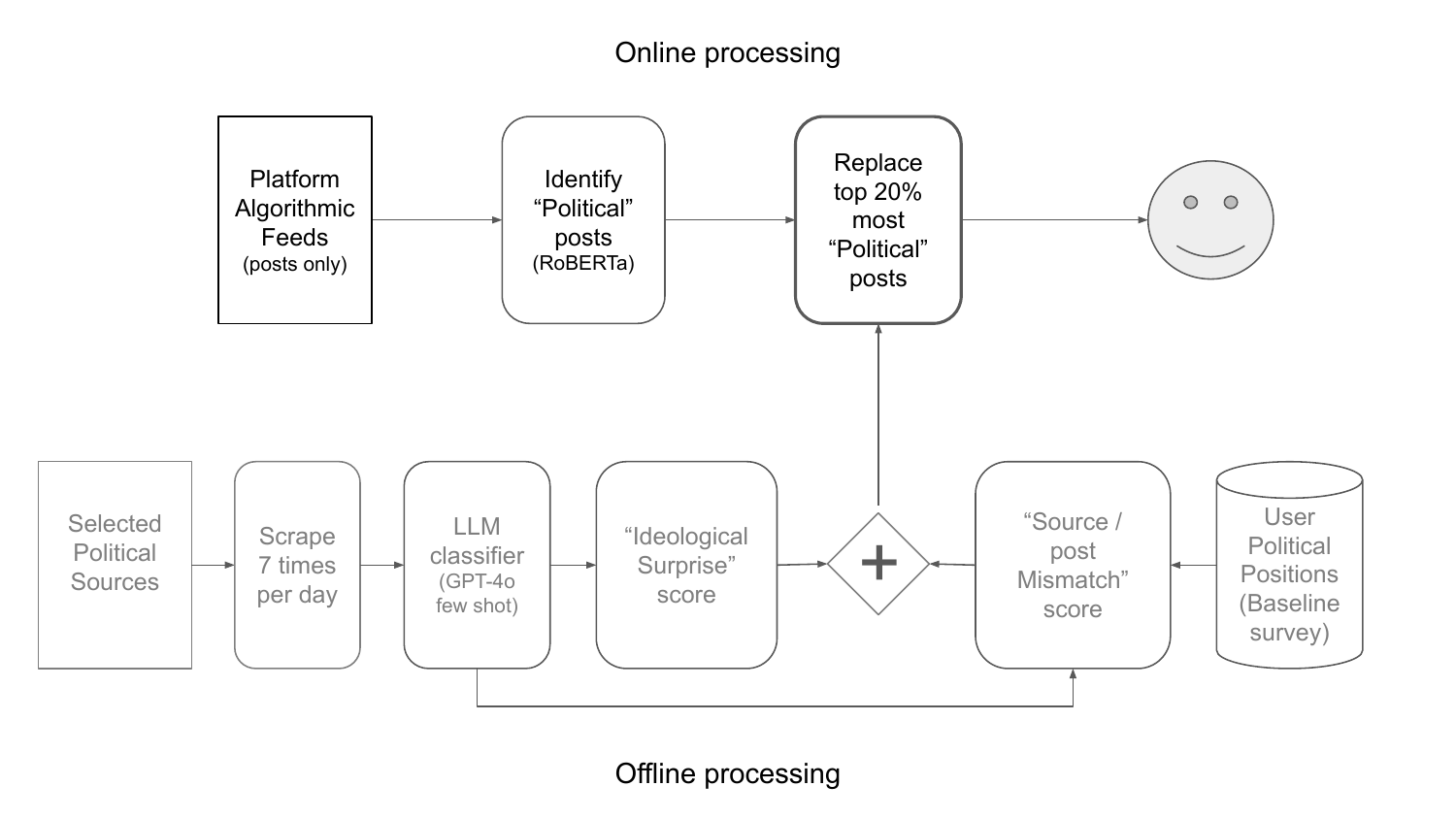}
    \caption{\textbf{Design of the {\surprising} ranking algorithm.}}
    \label{fig:perspective_design}
\end{figure}

\begin{figure}
    \centering
    \includegraphics[width=\linewidth]{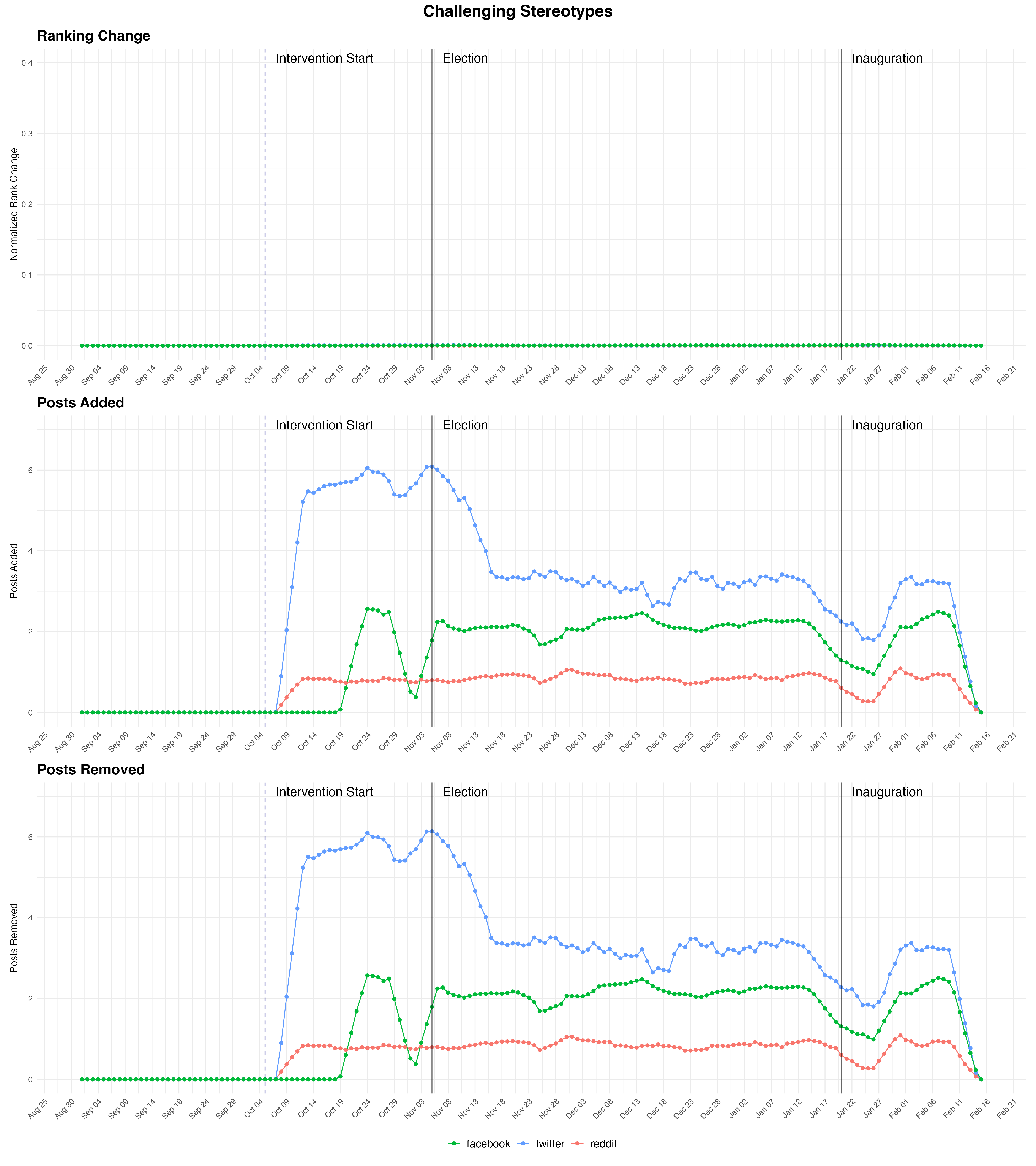}
    \caption{\textbf{Timeline of post rank change, posts added, and posts deleted for \emph{{\surprising}}}, 5 day moving average. Normalized rank change is the average numerical rank change for each item, divided by the length of the item slate.}
    \label{fig:surprising_timeline}
\end{figure}

\begin{figure}
    \centering
    \includegraphics[width=\linewidth]{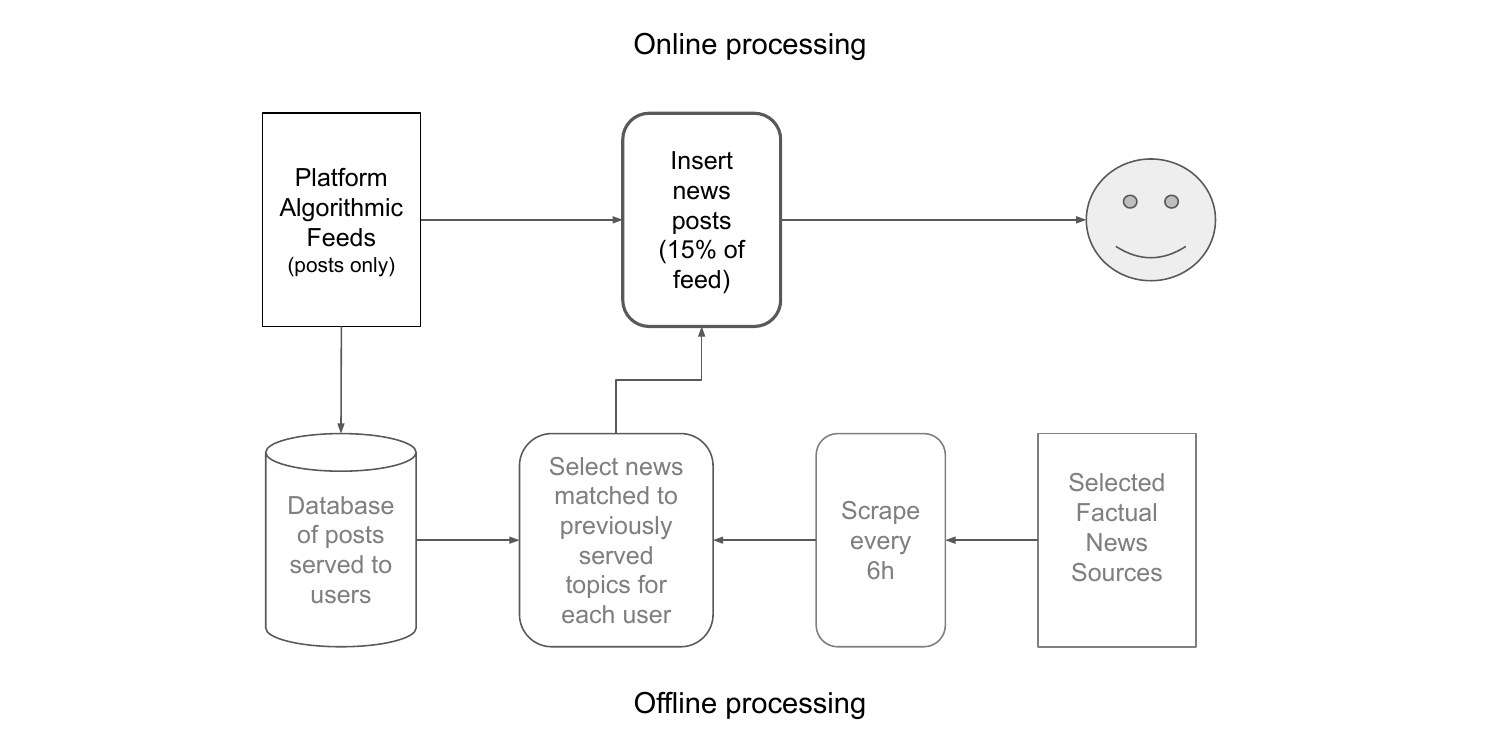}
    \caption{\textbf{Design of the {\addnews} ranking algorithm.}}
    \label{fig:perspective_design}
\end{figure}

\begin{figure}
    \centering
    \includegraphics[width=\linewidth]{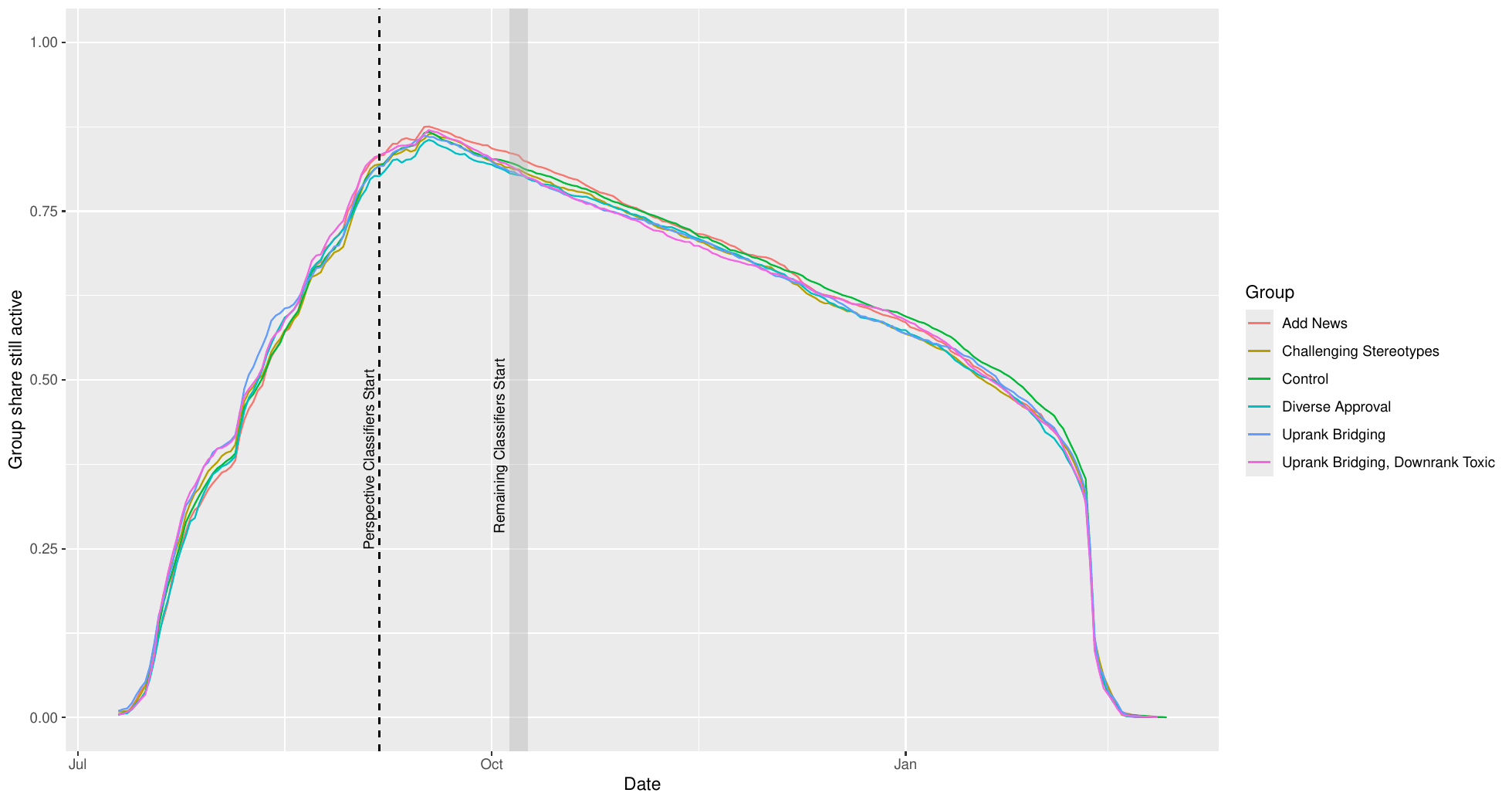}
    \caption{\textbf{The share of individuals in each group over time}, out of the total number ever validly enrolled in our sample (i.e. completed the baseline survey and kept the extension installed for at least one week). Recruitment ran until late September, which is why all shares increase until that point. Attrition is fairly steady for all groups. Note that we began automatically removing the extension in February of 2025, causing the sharp drop at that point.}
    \label{fig:survival_over_time}
\end{figure}


\clearpage

\begin{table}
    \centering
    \caption{\textbf{Results on all preregistered survey outcomes, estimated separately for each treatment arm.} All survey measures normalized using the control group mean and SD, except for Trust which is on a 0-100 scale. Bridging Pooled combines \textit{{\perspectivearm}} and \textit{{\perspectivearmdown}}. (* $p<0.1$, ** $p<0.05$, *** $p<0.01$)}
    \vspace{1em}
    \resizebox{\textwidth}{!}{
        \begin{tabular}{lllllllll}
\hline
& lhs: Polarization & lhs: Neely & lhs: Empathy & lhs: Wellbeing & lhs: Violence & lhs: Perceptions & lhs: Knowledge & lhs: Trust \\ \hline
Uprank Bridging, Downrank Toxic &&&&&&&& \\
Treated & -0.042** & -0.016 & 0.032 & 0.003 & 0.014 & -0.017 & -0.028 & 0.097** \\
& (0.017) & (0.019) & (0.041) & (0.027) & (0.026) & (0.040) & (0.066) & (0.041) \\
N & 5361 & 5318 & 3723 & 5360 & 5361 & 5361 & 5361 & 5361 \\
Uprank Bridging &&&&&&&& \\
Treated & -0.015 & -0.025 & -0.004 & -0.006 & -0.006 & 0.011 & 0.041 & 0.025 \\
& (0.019) & (0.019) & (0.043) & (0.028) & (0.027) & (0.040) & (0.067) & (0.042) \\
N & 5335 & 5288 & 3715 & 5335 & 5335 & 5335 & 5334 & 5335 \\
Challenging Stereotypes &&&&&&&& \\
Treated & -0.020 & -0.064*** & 0.027 & 0.013 & 0.030 & 0.072* & -0.019 & 0.041 \\
& (0.019) & (0.017) & (0.039) & (0.028) & (0.028) & (0.037) & (0.066) & (0.041) \\
N & 5523 & 5481 & 3845 & 5522 & 5523 & 5523 & 5522 & 5523 \\
Add News &&&&&&&& \\
Treated & -0.044** & -0.068*** & -0.049 & -0.016 & 0.036 & 0.024 & -0.089 & 0.068 \\
& (0.017) & (0.018) & (0.039) & (0.028) & (0.027) & (0.041) & (0.066) & (0.042) \\
N & 5482 & 5431 & 3779 & 5482 & 5482 & 5482 & 5480 & 5482 \\
Diverse Approval &&&&&&&& \\
Treated & -0.011 & -0.010 & 0.058 & -0.013 & 0.021 & -0.004 & -0.019 & 0.078* \\
& (0.018) & (0.018) & (0.042) & (0.028) & (0.025) & (0.040) & (0.067) & (0.043) \\
N & 5353 & 5310 & 3694 & 5352 & 5353 & 5353 & 5353 & 5353 \\
Bridging Pooled &&&&&&&& \\
Treated & -0.029* & -0.021 & 0.013 & -0.001 & 0.004 & -0.003 & 0.007 & 0.062* \\
& (0.015) & (0.015) & (0.034) & (0.023) & (0.021) & (0.033) & (0.055) & (0.034) \\
N & 7115 & 7056 & 4981 & 7114 & 7115 & 7115 & 7114 & 7115 \\
\hline
\end{tabular}

    }
    \label{tab:results_survey_primary}
\end{table}

\begin{table}
    \centering
    \caption{\textbf{Results on all preregistered survey outcomes, aggregated across all treatment arms.} All survey measures normalized using the control group mean and SD (* $p<0.1$, ** $p<0.05$, *** $p<0.01$)}
    \vspace{1em}
    \resizebox{\textwidth}{!}{
        \begin{tabular}{lllllllll}
\hline
& lhs: Polarization & lhs: Neely & lhs: Empathy & lhs: Wellbeing & lhs: Violence & lhs: Perceptions & lhs: Knowledge & lhs: Trust \\ \hline
Treated & \num{-0.027}** & \num{-0.038}*** & \num{0.012} & \num{-0.003} & \num{0.020} & \num{0.018} & \num{-0.024} & \num{0.062}** \\
& (\num{0.012}) & (\num{0.012}) & (\num{0.028}) & (\num{0.019}) & (\num{0.018}) & (\num{0.027}) & (\num{0.046}) & (\num{0.029}) \\
N & \num{12730} & \num{12628} & \num{8928} & \num{12727} & \num{12730} & \num{12730} & \num{12726} & \num{12730} \\
\hline
\end{tabular}

    }
    \label{tab:results_survey_primary_pooled}
\end{table}

\begin{table}[]
    \centering\
    \caption{\textbf{Heterogeneous effects on all preregistered survey outcomes, aggregated across all treatment arms} (* $p<0.1$, ** $p<0.05$, *** $p<0.01$)}
    \vspace{1em}
    \resizebox{\textwidth}{!}{
    \begin{tabular}{lllllllll}
\hline
& lhs: Polarization & lhs: Neely & lhs: Empathy & lhs: Wellbeing & lhs: Violence & lhs: Perceptions & lhs: Knowledge & lhs: Trust \\ \hline
Democrat &&&&&&&& \\
Treated & -0.032** & -0.015 & 0.048 & 0.012 & 0.024 & 0.062* & -0.147*** & 0.075** \\
& (0.015) & (0.015) & (0.034) & (0.023) & (0.022) & (0.033) & (0.057) & (0.036) \\
Treated:DemocratTRUE & 0.006 & -0.035** & -0.056* & -0.027 & -0.004 & -0.073** & 0.194*** & -0.020 \\
& (0.013) & (0.014) & (0.031) & (0.021) & (0.020) & (0.030) & (0.050) & (0.032) \\
$F: \beta_{\rm{Treated}} + \beta_{\rm{\times}} = 0$ & 3.54* & 13.95*** & 0.08 & 0.51 & 1.02 & 0.15 & 0.90 & 3.24* \\
N & 12696 & 12597 & 8899 & 12693 & 12696 & 12696 & 12692 & 12696 \\
White &&&&&&&& \\
Treated & -0.026 & -0.049*** & 0.049 & 0.018 & 0.006 & -0.032 & -0.108* & 0.107*** \\
& (0.016) & (0.015) & (0.035) & (0.025) & (0.026) & (0.035) & (0.058) & (0.036) \\
Treated:WhiteTRUE & 0.000 & 0.017 & -0.056* & -0.030 & 0.021 & 0.072** & 0.127** & -0.067** \\
& (0.015) & (0.014) & (0.032) & (0.022) & (0.023) & (0.032) & (0.051) & (0.032) \\
$F: \beta_{\rm{Treated}} + \beta_{\rm{\times}} = 0$ & 4.00** & 5.87** & 0.06 & 0.41 & 2.14 & 1.97 & 0.14 & 1.78 \\
N & 12708 & 12607 & 8915 & 12705 & 12708 & 12708 & 12704 & 12708 \\
Male &&&&&&&& \\
Treated & -0.032** & -0.040*** & 0.009 & -0.018 & 0.016 & 0.001 & -0.074 & 0.078** \\
& (0.014) & (0.014) & (0.030) & (0.021) & (0.019) & (0.030) & (0.051) & (0.031) \\
Treated:MaleTRUE & 0.012 & 0.004 & 0.008 & 0.031 & 0.008 & 0.037 & 0.107** & -0.036 \\
& (0.013) & (0.013) & (0.030) & (0.020) & (0.020) & (0.029) & (0.048) & (0.030) \\
$F: \beta_{\rm{Treated}} + \beta_{\rm{\times}} = 0$ & 1.98 & 6.38** & 0.24 & 0.39 & 1.23 & 1.50 & 0.40 & 1.59 \\
N & 12728 & 12626 & 8926 & 12725 & 12728 & 12728 & 12724 & 12728 \\
\hline
\end{tabular}

    }
    \label{tab:results_survey_het}
\end{table}

\begin{table}[]
    \centering
    \caption{\textbf{Effects on the individual components of our polarization index, pooled across arms} Thermometer is the difference between inparty and outparty affect in 0-100 scale units (degrees), Comfortable Friends is in 1-7 Likert scale units, all other columns are standardized effects. Outparty Affect combines outparty feeling thermometer and comfort with outparty friends, Inparty Affect combines combines inparty feeling thermometer and comfort with inparty friends (* $p<0.1$, ** $p<0.05$, *** $p<0.01$).}
    \vspace{1em}
    \footnotesize
    \begin{tabular}{lllll}
\hline
& Thermometer & Comfortable Friends & Outparty Affect & Inparty Affect \\ \hline
Treated & \num{-1.480}* & \num{-0.065} & \num{0.025} & \num{-0.028} \\
& (\num{0.786}) & (\num{0.042}) & (\num{0.017}) & (\num{0.020}) \\
Mean & \num{40.44} & \num{1.49} & \num{-0.091} & \num{-0.017} \\
SD & \num{35.07} & \num{1.89} & \num{0.837} & \num{0.843} \\
N & \num{12730} & \num{12730} & \num{12730} & \num{12730} \\
\hline
\end{tabular}

    \label{tab:results_survey_components}
\end{table}

\begin{table}
    \centering
    \caption{\textbf{Main treatment effects on active time and total engagements by platform and combined} (* $p<0.1$, ** $p<0.05$, *** $p<0.01$) }
    \vspace{1em}
    \footnotesize
    \begin{tabular}{lllll}
\hline
& Facebook & X/Twitter & Reddit & CombinedFXR \\ \hline
Active Time &&&& \\
Treated & -0.368*** & 0.319*** & -0.195*** & -0.186* \\
& (0.131) & (0.078) & (0.072) & (0.108) \\
N & 1636498 & 1636498 & 1636498 & 1636498 \\
Total Engagement &&&& \\
Treated & -0.242** & 0.413*** & -0.040 & 0.131 \\
& (0.101) & (0.092) & (0.042) & (0.120) \\
N & 1636498 & 1636498 & 1636498 & 1636498 \\
\hline
\end{tabular}

\end{table}

\begin{table}
    \centering
    \caption{\textbf{Active Time treatment effects by platform and individual ranker arm} (* $p<0.1$, ** $p<0.05$, *** $p<0.01$)}
    \vspace{1em}
    \scriptsize
    \begin{tabular}{lllll}
\hline
& Facebook & X/Twitter & Reddit & CombinedFXR \\ \hline
Uprank Bridging, Downrank Toxic &&&& \\
Treated & -0.366*** & 0.477*** & -0.221*** & -0.059 \\
& (0.119) & (0.086) & (0.066) & (0.145) \\
Mean & 5.58 & 2.96 & 1.81 & 8.55 \\
SD & 20.26 & 19.74 & 11.02 & 28.93 \\
N & 696965 & 696965 & 696965 & 696965 \\
Uprank Bridging &&&& \\
Treated & -0.302* & 0.151* & -0.191* & -0.258* \\
& (0.173) & (0.089) & (0.115) & (0.150) \\
Mean & 5.47 & 2.98 & 1.79 & 8.49 \\
SD & 19.36 & 19.89 & 10.99 & 28.51 \\
N & 682210 & 682210 & 682210 & 682210 \\
Challenging Stereotypes &&&& \\
Treated & -0.458*** & 0.355*** & -0.144* & 0.064 \\
& (0.086) & (0.084) & (0.082) & (0.120) \\
Mean & 5.55 & 2.89 & 1.79 & 8.43 \\
SD & 19.57 & 19.39 & 12.72 & 28.92 \\
N & 701306 & 701306 & 701306 & 701306 \\
Add News &&&& \\
Treated & -0.380*** & 0.187* & -0.036 & -0.032 \\
& (0.095) & (0.095) & (0.060) & (0.169) \\
Mean & 5.47 & 2.95 & 1.73 & 8.42 \\
SD & 19.55 & 19.73 & 10.62 & 28.44 \\
N & 702968 & 702968 & 702968 & 702968 \\
Diverse Approval &&&& \\
Treated & 0.064 & -0.505*** & 0.011 & -0.303** \\
& (0.085) & (0.077) & (0.040) & (0.144) \\
Mean & 5.47 & 3.02 & 1.71 & 8.45 \\
SD & 19.41 & 19.98 & 10.41 & 28.42 \\
N & 693773 & 693773 & 693773 & 693773 \\
Bridging Pooled &&&& \\
Treated & -0.369*** & 0.319*** & -0.195*** & -0.186* \\
& (0.120) & (0.075) & (0.068) & (0.105) \\
Mean & 5.64 & 2.98 & 1.83 & 8.66 \\
SD & 20.26 & 19.66 & 11.29 & 29.03 \\
N & 918994 & 918994 & 918994 & 918994 \\
\hline
\end{tabular}

\end{table}

\begin{table}
    \centering
    \caption{\textbf{Total Engagement treatment effects by platform and individual ranker arm} (* $p<0.1$, ** $p<0.05$, *** $p<0.01$)}
    \vspace{1em}
    \scriptsize
    \begin{tabular}{lllll}
\hline
& Facebook & X/Twitter & Reddit & CombinedFXR \\ \hline
Uprank Bridging, Downrank Toxic &&&& \\
Treated & -0.348*** & 0.479*** & 0.062** & 0.193 \\
& (0.111) & (0.097) & (0.027) & (0.143) \\
Mean & 3.72 & 1.99 & 0.273 & 5.99 \\
SD & 20.48 & 18.16 & 4.98 & 28.26 \\
N & 696965 & 696965 & 696965 & 696965 \\
Uprank Bridging &&&& \\
Treated & -0.098 & 0.292*** & -0.148** & 0.046 \\
& (0.123) & (0.103) & (0.060) & (0.158) \\
Mean & 3.62 & 2.04 & 0.30 & 5.96 \\
SD & 19.56 & 18.62 & 5.52 & 28.07 \\
N & 682210 & 682210 & 682210 & 682210 \\
Challenging Stereotypes &&&& \\
Treated & -0.278** & 0.756*** & 0.157*** & 0.635*** \\
& (0.117) & (0.109) & (0.024) & (0.168) \\
Mean & 3.66 & 1.95 & 0.266 & 5.87 \\
SD & 19.79 & 17.73 & 4.85 & 27.46 \\
N & 701306 & 701306 & 701306 & 701306 \\
Add News &&&& \\
Treated & -0.066 & -0.033 & 0.017 & -0.082 \\
& (0.109) & (0.108) & (0.026) & (0.150) \\
Mean & 3.58 & 2.11 & 0.304 & 5.99 \\
SD & 19.42 & 20.23 & 5.41 & 29.08 \\
N & 702968 & 702968 & 702968 & 702968 \\
Diverse Approval &&&& \\
Treated & -0.162 & -0.495*** & 0.105*** & -0.552*** \\
& (0.115) & (0.102) & (0.020) & (0.171) \\
Mean & 3.62 & 2.06 & 0.263 & 5.95 \\
SD & 19.90 & 19.51 & 4.84 & 28.74 \\
N & 693773 & 693773 & 693773 & 693773 \\
Bridging Pooled &&&& \\
Treated & -0.242** & 0.413*** & -0.040 & 0.131 \\
& (0.096) & (0.087) & (0.040) & (0.122) \\
Mean & 3.83 & 2.02 & 0.287 & 6.13 \\
SD & 20.69 & 18.35 & 5.35 & 28.65 \\
N & 918994 & 918994 & 918994 & 918994 \\
\hline
\end{tabular}

\end{table}

\begin{table}[]
    \centering
    \caption{\textbf{Heterogeneous effects on active time, aggregated across all treatment arms} (* $p<0.1$, ** $p<0.05$, *** $p<0.01$)}
    \vspace{1em}
    \footnotesize
    \begin{tabular}{lllll}
\hline
& Facebook & X/Twitter & Reddit & CombinedFXR \\ \hline
Democrat &&&& \\
Treated & -0.731*** & 0.359*** & -0.383*** & -0.692*** \\
& (0.127) & (0.125) & (0.087) & (0.212) \\
Treated:Democrat & 0.498* & -0.089 & 0.304** & 0.722* \\
& (0.259) & (0.153) & (0.142) & (0.370) \\
N & 1577896 & 1577896 & 1577896 & 1577896 \\
White &&&& \\
Treated & -0.177 & 0.098 & -0.299*** & 0.026 \\
& (0.163) & (0.157) & (0.079) & (0.217) \\
Treated:White & -0.318* & 0.321* & 0.191 & -0.396 \\
& (0.181) & (0.181) & (0.128) & (0.355) \\
N & 1581238 & 1581238 & 1581238 & 1581238 \\
Male &&&& \\
Treated & -0.294 & 0.255*** & -0.380*** & -0.729*** \\
& (0.186) & (0.079) & (0.100) & (0.167) \\
Treated:Male & -0.256 & 0.088 & 0.392*** & 1.049*** \\
& (0.222) & (0.174) & (0.121) & (0.325) \\
N & 1583807 & 1583807 & 1583807 & 1583807 \\
\hline
\end{tabular}

    \label{tab:results_platform_het}
\end{table}

\begin{table}[]
    \centering
    \caption{\textbf{Heterogeneous effects on total engagements aggregated across all treatment arms} (* $p<0.1$, ** $p<0.05$, *** $p<0.01$)}
    \vspace{1em}
    \footnotesize
    \begin{tabular}{lllll}
\hline
& Facebook & X/Twitter & Reddit & CombinedFXR \\ \hline
Democrat &&&& \\
Treated & -0.438* & 0.252* & -0.180** & -0.366 \\
& (0.225) & (0.145) & (0.072) & (0.248) \\
Treated:Democrat & 0.264 & 0.231 & 0.221*** & 0.715* \\
& (0.407) & (0.166) & (0.065) & (0.374) \\
N & 1577896 & 1577896 & 1577896 & 1577896 \\
White &&&& \\
Treated & -0.259** & 0.167 & -0.022 & -0.113 \\
& (0.125) & (0.149) & (0.019) & (0.158) \\
Treated:White & -0.021 & 0.396** & -0.016 & 0.359 \\
& (0.199) & (0.193) & (0.059) & (0.277) \\
N & 1581238 & 1581238 & 1581238 & 1581238 \\
Male &&&& \\
Treated & 0.266 & 0.121 & -0.194*** & 0.193 \\
& (0.171) & (0.093) & (0.069) & (0.187) \\
Treated:Male & -1.185*** & 0.542*** & 0.345*** & -0.298 \\
& (0.251) & (0.168) & (0.059) & (0.297) \\
N & 1583807 & 1583807 & 1583807 & 1583807 \\
\hline
\end{tabular}

    \label{tab:results_platform_het_eng}
\end{table}

\begin{table}[]
    \centering
    \caption{\textbf{Outcomes for in-feed survey questions.} (* $p<0.1$, ** $p<0.05$, *** $p<0.01$)}
    \vspace{1em}
    \resizebox{\textwidth}{!}{
        \begin{tabular}{llllll}
\hline
& OutpartyFeeling & CampaignInterest & FeedSentiment & PerceivedViolenceLikelihood & PerspectiveDifficulty \\ \hline
Treatment & -0.469 & 0.128 & -0.016 & 0.543 & -0.067 \\
& (0.463) & (0.414) & (0.282) & (0.496) & (0.342) \\
N & 2721 & 1614 & 2297 & 1821 & 2379 \\
\hline
\end{tabular}

    }
    \label{tab:in_feed}
\end{table}

\begin{table}[h]
\centering
    \caption{\textbf{Number of items re-ranked on each page, by platform and content type.}}
    \vspace{1em}
    \footnotesize
    \begin{tabular}{lll}
        \hline
        Platform & Content & Slate size \\
        \hline
        Reddit   & Posts    & 50 \\
        Reddit   & Comments & all available \\
        Twitter  & Posts    & 50 \\
        Twitter  & Replies  & 50 (or all available) \\
        Facebook & Posts    & 10 (first pass), then 35 (second pass) \\
        Facebook & Comments & 50 (or all available) \\
        \hline
    \end{tabular}
    \label{tab:slate_length}
\end{table}

\begin{table}
    \centering
    \caption{\textbf{Experimental attributes from Perspective API.} AUC measures performance on the tests data sets in \cite{schmer-galunder_annotator_2024}.}
    \vspace{1em}
    \footnotesize
    \begin{tabular}{|p{2.5cm}|p{10cm}|p{1cm}|}
        \hline
        \textbf{Attribute} & \textbf{Definition} & \textbf{AUC-ROC} \\
        \hline
        Affinity & References shared interests, motivations or outlooks between the comment author and another individual, group or entity. & 0.86 \\
        \hline
        Compassion & Identifies with or shows concern, empathy, or support for the feelings/emotions of others. & 0.91 \\
        \hline
        Curiosity & Attempts to clarify or ask follow-up questions to better understand another person or idea. & 0.87 \\
        \hline
        Nuance & Incorporates multiple points of view in an attempt to provide a full picture or contribute useful detail and/or context. & 0.88 \\
        \hline
        Personal story & Includes a personal experience or story as a source of support for the statements made in the comment. & 0.95 \\
        \hline
        Reasoning & Makes specific or well-reasoned points to provide a fuller understanding of the topic without disrespect or provocation. & 0.87 \\
        \hline
        Respect & Shows deference or appreciation to others, or acknowledges the validity of another person. & 0.96 \\
        \hline
        Alienation & Portrays someone as inferior, implies a lack of belonging, or frames the statement in an us vs. them context. & 0.93 \\
        \hline
        Fearmongering & Deliberately arouses fear or alarm about a particular issue. & 0.93 \\
        \hline
        Stereotyping & Asserts something to be true for either of all members of a certain group or of an indefinite part of that group. & 0.88 \\
        \hline
        Moral outrage & Anger, disgust, or frustration directed toward other people or entities who seem to violate the author’s ethical values or standards. & 0.93 \\
        \hline
        Scapegoating & Blames a person or entity for the wrongdoings, mistakes, or faults of others, especially for reasons of expediency. & 0.91 \\
        \hline
        Identity attack & Negative or hateful comments targeting someone because of their identity. & 0.97 \\
        \hline
        Insult & Insulting, inflammatory, or negative comment towards a person or a group of people. & 0.97 \\
        \hline
    \end{tabular}
    \label{tab:perspective_attributes}
\end{table}

\begin{table}
\centering
\caption{\textbf{Classifier weights for {\perspectivearm} ranker}}
\vspace{1em}
\footnotesize
\begin{tabular}{|p{2.5cm}|p{1.5cm}|}
\hline
\textbf{Attribute} & \textbf{Weight} \\
\hline
Reasoning & 1/6 \\
\hline
Personal story & 1/6 \\
\hline
Affinity & 1/6 \\
\hline
Compassion & 1/6 \\
\hline
Respect & 1/6 \\
\hline
Curiosity & 1/6 \\
\hline
\end{tabular}
\label{tab:perspective_uprank_weights}
\end{table}

\begin{table}
\centering
\caption{\textbf{Classifier weights for {\perspectivearmdown}}}
\vspace{1em}
\footnotesize
\begin{tabular}{|p{2.5cm}|p{1.5cm}|}
\hline
\textbf{Attribute} & \textbf{Weight} \\
\hline
Reasoning & 1/6 \\
\hline
Personal story & 1/6 \\
\hline
Affinity & 1/6 \\
\hline
Compassion & 1/6 \\
\hline
Respect & 1/6 \\
\hline
Curiosity & 1/6 \\
\hline
Fearmongering & -1/6 \\
\hline
Stereotyping & -1/6 \\
\hline
Scapegoating & -1/18 \\
\hline
Moral outrage & -1/18 \\
\hline
Alienation & -1/18 \\
\hline
Toxicity & -1/8 \\
\hline
Identity attack & -1/8 \\
\hline
Insult & -1/8 \\
\hline
Threat & -1/8 \\
\hline
\end{tabular}
\label{tab:perspective_uprank_downrank_weights}
\end{table}

\begin{table}[ht]
\centering
\caption{\textbf{Participant recruitment companies, recruitment funnel stages, and conversion rates.} ``Screened in'' counts people eligible to participate (see table \ref{tab:screener}). Forthright and Symmetric implemented their own screening survey. Facebook and Twitter refer to platform ads. We were unable to record recruitment channel information for some users due to ad blockers and other technical issues.}
\vspace{1em}
\footnotesize
\setlength{\tabcolsep}{4pt}
\resizebox{\textwidth}{!}{
    \begin{tabular}{lccccccc}
    \hline
    \textbf{Source} & \textbf{Started screening} & \textbf{Conv\%} & \textbf{Screened in} & \textbf{Conv\%} & \textbf{Installed Extension} & \textbf{Conv\%} & \textbf{Kept installed one week} \\
    \hline
    Rep Data & 69,854 & 32 & 22,615 & 19 & 4,374 & 53 & 2,338 \\
    Audience Align & 56,167 & 41 & 22,778 & 11 & 2,583 & 44 & 1,131 \\
    Pure Spectrum & 54,392 & 33 & 18,117 & 18 & 3,321 & 51 & 1,705 \\
    Eleven MR & 52,835 & 40 & 21,317 & 11 & 2,248 & 69 & 1,554 \\
    Prodege & 6,967 & 15 & 1,068 & 25 & 266 & 80 & 213 \\
    Cloud Research & 4,516 & 37 & 1,678 & 66 & 1,102 & 74 & 811 \\
    Sago & 2,630 & 39 & 1,032 & 11 & 115 & 49 & 56 \\
    Cint & 2,564 & 27 & 683 & 9 & 64 & 45 & 29 \\
    Roots & 2,092 & 19 & 388 & 22 & 84 & 69 & 58 \\
    Innovate MR & 1,634 & 44 & 723 & 23 & 167 & 41 & 69 \\
    Positly & 1,328 & 14 & 188 & 40 & 76 & 71 & 54 \\
    Borderless & 470 & 24 & 113 & 28 & 32 & 56 & 18 \\
    Forthright &  &  &  &  & 2,064 & 63 & 1,295 \\
    Symmetric &  &  &  &  & 155 & 70 & 109 \\
    Facebook &  &  &  &  & 38 & 82 & 31 \\
    Twitter &  &  &  &  & 25 & 84 & 21 \\
    Unknown &  &  &  &  & 374 & 56 & 209 \\
    \hline
    \textbf{Total} & \textbf{255,449} & \textbf{36} & \textbf{90,700} & \textbf{19} & \textbf{17,088} & \textbf{57} & \textbf{9,701} \\
    \hline
    \end{tabular}
}
\label{tab:recruitment_sources}
\end{table}

\begin{table}[ht]
\centering
\caption{\textbf{Recruitment screening survey}}
\vspace{1em}
\footnotesize
\begin{tabular}{p{0.6\textwidth} p{0.3\textwidth}}
\hline
\textbf{Question} & \textbf{Passing answer}  \\
\hline
What device are you currently using? & Laptop or Desktop \\
\hline
Are you currently using the Chrome browser on your computer? & Yes \\
\hline
Which of the following social media apps/websites do you use? & At least one of Facebook, Twitter, Reddit \\
\hline
What percentage of the time would you say you use social media from your desktop computer (including laptops), as opposed to your phone or mobile device? & $\geq 30$ percent  \\
\hline
We are recruiting people to participate in an academic research study that involves installing a browser extension. You will be paid a total of $\$10$ for your participation in this study, plus a chance to win $\$500$, if you keep the extension installed for five months. & Yes \\
This is a study run by researchers at UC Berkeley who are studying the effects of social media and trying to make a better user experience. We would like to collect anonymized data about your online experiences for research purposes only, and make small changes to which content you see on social media. You will be paid $\$5$ for each of two surveys over several months. \\
Are you interested in participating?  \\
\hline
\end{tabular}
\label{tab:screener}
\end{table}

\begin{table}[h]
    \centering
    \caption{\textbf{Participant sample descriptive statistics}}
    \vspace{1em}
    \footnotesize    
    
\begin{tabular}[t]{l|r|r|r|r|r|r}
\hline
Variable & Mean & Median & SD & Min & Max & N\\
\hline
Bachelors & 0.435 & 0.000 & 0.496 & 0 & 1.000 & 9386\\
\hline
BaselineTimeFacebook & 11.151 & 2.594 & 23.258 & 0 & 435.676 & 6002\\
\hline
BaselineTimeReddit & 4.819 & 0.762 & 15.999 & 0 & 396.711 & 4882\\
\hline
BaselineTimeX/Twitter & 5.839 & 0.214 & 22.639 & 0 & 365.978 & 7014\\
\hline
Democrat & 0.616 & 1.000 & 0.486 & 0 & 1.000 & 9346\\
\hline
ExtensionData & 1.000 & 1.000 & 0.000 & 1 & 1.000 & 9709\\
\hline
IncomeK & 68.588 & 62.500 & 47.406 & 10 & 175.000 & 9323\\
\hline
Male & 0.471 & 0.000 & 0.499 & 0 & 1.000 & 9383\\
\hline
SocialMediaMoreThan90Min & 0.517 & 1.000 & 0.500 & 0 & 1.000 & 9386\\
\hline
White & 0.649 & 1.000 & 0.477 & 0 & 1.000 & 9368\\
\hline
Young & 0.326 & 0.000 & 0.469 & 0 & 1.000 & 9386\\
\hline
\end{tabular}

    \label{tab:descriptives}
\end{table}

\begin{table}[h]
    \centering
    \caption{\textbf{Balance Table}}
    \vspace{1em}
    \footnotesize
    \resizebox{\textwidth}{!}{
        \begin{tabular}{lllllll}
\hline
& Control (N=2729) &  & Prosocial Ranking (N=6980) &  &  &  \\ \cline{2-3}\cline{4-5}
& Mean & Std. Dev. & Mean & Std. Dev. & Diff. in Means & p \\ \hline
Male & 0.5 & 0.5 & 0.5 & 0.5 & -0.0 & 0.648 \\
White & 0.6 & 0.5 & 0.7 & 0.5 & 0.0 & 0.559 \\
Young & 0.3 & 0.5 & 0.3 & 0.5 & -0.0 & 0.622 \\
Bachelors & 0.4 & 0.5 & 0.4 & 0.5 & 0.0 & 0.882 \\
Democrat & 0.6 & 0.5 & 0.6 & 0.5 & -0.0 & 0.187 \\
IncomeK & 69.0 & 47.5 & 68.4 & 47.4 & -0.6 & 0.576 \\
SocialMediaMoreThan90Min & 0.5 & 0.5 & 0.5 & 0.5 & -0.0 & 0.088 \\
BaselineTimeFacebook & 10.4 & 21.2 & 11.4 & 24.0 & 1.1 & 0.099 \\
BaselineTimeTwitter & 6.5 & 24.7 & 5.6 & 21.8 & -0.9 & 0.141 \\
BaselineTimeReddit & 5.0 & 13.2 & 4.7 & 17.0 & -0.3 & 0.542 \\
\hline
\end{tabular}

    }
\end{table}

\begin{table}[]
    \centering
    \caption{\textbf{Regression models testing for differential survey attrition by treatment group} (* $p<0.1$, ** $p<0.05$, *** $p<0.01$) }
    \vspace{1em}
    \footnotesize
    \begin{tabular}{llll}
\hline
& (1) & (2) & (3) \\ \hline
(Intercept) & 0.534*** & -0.030 & 0.534*** \\
& (0.010) & (0.064) & (0.010) \\
Treatment & 0.003 & 0.021 &  \\
& (0.012) & (0.075) &  \\
SurvivalScore &  & 1.050*** &  \\
&  & (0.118) &  \\
Treatment:SurvivalScore &  & -0.031 &  \\
&  & (0.139) &  \\
GroupAN:Treatment &  &  & 0.007 \\
&  &  & (0.017) \\
GroupCS:Treatment &  &  & 0.022 \\
&  &  & (0.017) \\
GroupDA:Treatment &  &  & -0.015 \\
&  &  & (0.017) \\
GroupUB:Treatment &  &  & 0.019 \\
&  &  & (0.017) \\
GroupUBDT:Treatment &  &  & -0.016 \\
&  &  & (0.017) \\
N & 9269 & 9269 & 9269 \\
\hline
\end{tabular}

    \label{tab:survey_attrition}
\end{table}

\begin{table}
    \centering
    \caption{\textbf{Robustness to sample inclusion criteria.} The main specification includes participants who remained in the study for at least one week. This table replicates the main results using alternative minimum participation cutoffs of 2 days and 2 weeks. All survey measures normalized using the control group mean and SD. Active time outcomes are in relative terms (minutes per day per baseline hour) (* $p<0.1$, ** $p<0.05$, *** $p<0.01$).}
    \vspace{1em}
    \footnotesize
    \begin{tabular}{llllll}
\hline
& Polarization & Facebook & Twitter & Reddit & CombinedFTR \\ \hline
2-day cutoff &&&&& \\
Treated & -0.027** & -0.375*** & 0.292*** & -0.185*** & -0.213** \\
& (0.012) & (0.129) & (0.082) & (0.068) & (0.107) \\
N & 12815 & 1698840 & 1698840 & 1698840 & 1698840 \\
2-week cutoff &&&&& \\
Treated & -0.025** & -0.345*** & 0.361*** & -0.182** & -0.106 \\
& (0.012) & (0.117) & (0.079) & (0.076) & (0.111) \\
N & 12646 & 1581860 & 1581860 & 1581860 & 1581860 \\
\hline
\end{tabular}

    \label{app:robustness_inclusion}
\end{table}

\begin{table}
    \centering
    \caption{\textbf{Robustness to Twitter usage in the sample.} ``No Twitter Sample" excludes all participants with any baseline Twitter usage. ``Low Twitter Sample" excludes participants with baseline Twitter usage exceeding 5 minutes per week. All survey measures normalized using the control group mean and SD. Active time outcomes are in relative terms (minutes per day per baseline hour) (* $p<0.1$, ** $p<0.05$, *** $p<0.01$).}
    \vspace{1em}
    \footnotesize
    \begin{tabular}{llllll}
\hline
& Polarization & Facebook & Twitter & Reddit & CombinedFTR \\ \hline
No Twitter Sample &&&&& \\
Treated & -0.015 & -0.022 & 0.005 & -0.106** & -0.202* \\
& (0.016) & (0.085) & (0.033) & (0.051) & (0.112) \\
N & 6908 & 1079407 & 1079407 & 1079407 & 1079407 \\
Low Twitter Sample &&&&& \\
Treated & -0.027** & -0.184* & -0.040 & -0.177*** & -0.425*** \\
& (0.014) & (0.108) & (0.029) & (0.039) & (0.106) \\
N & 10184 & 1471811 & 1471811 & 1471811 & 1471811 \\
\hline
\end{tabular}

    \label{app:robustness_twitter}
\end{table}



\end{document}